\documentclass[pdflatex,sn-mathphys-num]{sn-jnl}

\usepackage{graphicx}%
\usepackage{multirow}%
\usepackage{amsmath,amssymb,amsfonts}%
\usepackage{amsthm}%
\usepackage{mathrsfs}%
\usepackage[title]{appendix}%
\usepackage{xcolor}%
\usepackage{textcomp}%
\usepackage{manyfoot}%
\usepackage{booktabs}%
\usepackage{algorithm}%
\usepackage{algorithmicx}%
\usepackage{algpseudocode}%
\usepackage{listings}%
\usepackage{subfigure}

\theoremstyle{thmstyleone}%
\theoremstyle{thmstyletwo}%

\theoremstyle{thmstylethree}%

\begin{document}

\title[Article Title]{Progressive-Iterative Fairing of Curves and Surfaces with Localized Control Point Adjustment}


\author[1]{\fnm{Jia} \sur{Lin}}

\author*[1,2]{\fnm{Hongwei} \sur{Lin}}\email{hwlin@zju.edu.cn}

\author[3]{\fnm{Weixian} \sur{Huang}}

\author[1]{\fnm{Azan} \sur{Zhang}}

\affil*[1]{\orgdiv{School of Mathematical Sciences}, \orgname{Zhejiang University}, \orgaddress{\city{Hangzhou}, \postcode{310058}, \country{China}}}

\affil[2]{\orgdiv{State Key Laboratory of CAD \& CG}, \orgname{Zhejiang University}, \orgaddress{\city{Hangzhou}, \postcode{310058},\country{China}}}

\affil[3]{ \orgname{ZWSOFT}, \orgaddress{ \city{Guangzhou}, \postcode{510623}, \country{China}}}


\abstract{Curve and surface fairing is crucial in computer-aided geometric design, influencing product quality, physical performance, and aesthetics. Traditional methods often apply global modifications, lacking fine-grained control. This paper introduces a novel progressive-iterative fairing method based on control point adjustment. By assigning independent weights to each control point, our approach enables precise, localized shape adjustments. The method functions both globally and locally, allowing for comprehensive shape fairing and fine control over the fairing effect. Furthermore, this paper provides an automatic control point selection method to adjust shapes, thereby eliminating the reliance on manual interaction.  Numerical experiments demonstrate the efficiency and effectiveness of our approach.}





\maketitle

\section{Introduction}\label{sec1}

Curve and surface fairing originates from practical requirements in engineering design and manufacturing, with significant applications in fields such as automotive body modeling\cite{WESTGAARD2001619}, aircraft wing design \cite{SplineBased2005}, and ship hull formation \cite{Sarz2006AnOA}. As a key research focus in computer-aided geometric design, fairness critically influences product quality, physical performance, aesthetics, and other attributes. Over the years, extensive studies have been conducted by researchers worldwide, resulting in the development of numerous effective fairing algorithms.

When existing preliminary curve or surface modelsare used, such as those derived from early hand-drawings or parametric design, local shape defects may arise due to rapid iterations in early design stages or parameter deviations. These models may also fail to meet stricter fairness requirements in specific applications. Direct reconstruction from data points would discard prior design efforts and increase workload; therefore, a fairing method that enhances fairness through partial parameter adjustments, without altering the overall topological structure, is highly desirable. A common approach to fairing involves adjusting control points, with the objective of improving fairness while minimizing deviation from the original shape. This problem is frequently formulated as an energy optimization or constrained energy optimization task. However, such methods typically apply global modifications even for local irregularities and lack parameters for fine-grained fairness control. Furthermore, they often require solving linear or nonlinear systems of equations, which can be computationally inefficient for complex surfaces or those containing numerous anomalies.

In this study, a curve and surface fairing method based on PIA (Progressive-Iterative Approximation\cite{SSJD15103100222743,SJESAFC0B52EF118BE43FA8FEFC84DC20EC7}) is proposed to address the limitations of traditional approaches. Through iterative updates of control points, a series of progressively faired curves and surfaces are generated. In each iteration, offset vectors and fairing vectors are first constructed for each control point. A weighted sum of these vectors then serves as the differential update vector. New control points are obtained by adding these differential vectors to the current positions. Moreover, an independent weight is assigned to each control point, enabling fine-grained adjustment of geometric shape. This allows for localized fairness improvement while preserving important local features. The convergence of the proposed method is rigorously demonstrated, and it is shown that conventional energy-optimization fairing models constitute a special case of this framework. The main contributions of this paper are threefold:

\begin{itemize}
	\item A curve and surface fairing method is proposed based on the Progressive-Iterative Approximation (PIA) framework, which minimizes shape deviation during adjustment while avoiding the need to solve large systems of linear equations. This approach provides a flexible algorithm with high computational efficiency.
	\item This method functions both globally, enabling overall shape fairing, and locally, permitting fine control over the fairing effect through adjustable weights while preserving the original local features of the curve or surface.
	\item The proposed method is highly flexible, allowing attributes such as control point fairing weights to be modified during each iteration.
\end{itemize}

This paper presents an iterative fairing algorithm using B-spline curves and surfaces as illustrative examples. However, the proposed method is also applicable to other blended curves and surfaces, including B$\acute{e}$zier, NURBS, and T-spline representations.

The paper is organized as follows: Section 2 introduces the related work; Section 3 presents the curve and surface fairing algorithm and provides detailed descriptions; Section 4 analyzes the convergence of the proposed algorithm from a theoretical perspective; Section 5 verifies the effectiveness of the proposed algorithm through numerical experiments and provides a detailed analysis of the experimental results; Section 6 concludes the paper and discusses future research directions.

\section{Related work}

\subsection{Curve and surface fairing} 

The fairness of curves and surfaces represents a critical issue in Computer-Aided Geometric Design (CAGD). Issues such as precision loss and noise interference during data exchange between CAD systems make the fairing of existing geometric models both essential and practical.
Hosaka\cite{Hosaka1969} formulated the fairing problem as an energy minimization task, using a weighted objective function combining shape deviation and a fairness measure based on spline energy. Kjellander et al.\cite{KJELLANDER1983175} introduced an interactive fairing algorithm that improves curve fairness by modifying points where the displacement vector exceeds a predefined threshold, employing a node deletion and reinsertion strategy. While capable of global fairness improvement, this method suffers from slow computation when searching for optimal knot removal.
Farin et al.\cite{FARIN198791} presented a local fairing method for B-spline curves based on knot elimination and insertion to achieve a smoother curvature distribution. Lott and Pullin\cite{Lott1988} proposed an approach that automatically repositions surface control points via constrained minimization, using curvature as the objective function and shape deviation as the constraint.
Eck and Hadenfeld\cite{1995Localcurve, 1995Local}  developed a method to iteratively adjust a single B-spline control point to minimize the integral of the squared derivative of the curve or surface. Nishiyama et al.\cite{Nishiyama20027} introduced a technique for removing irregularities in B-spline surfaces by fairing circular highlight lines. Users first identify irregular regions interactively, which are then faired using cubic Hermite interpolation. The surface is subsequently modified by solving a nonlinear system to match the faired highlights, resulting in an improved surface. More recently, optimization-based methods\cite{ICWCAAP} have been proposed to construct Bézier interpolatory curves with parabolic-like curvature distribution. These methods employ tailored energy functionals to achieve $G^2$ smoothness, fairness, and strong locality through efficient local optimization.
Song et al.\cite{RTGSIFFSLSTT} propose a real-time global smoothing method for five-axis short line toolpaths. The lightweight two-layer structure generates $G^1$-continuous paths and $C^2$-continuous trajectories while maintaining high computational efficiency.
Li et al.\cite{AICCSAIF} propose an improved cubic cardinal spline method. It maintains traditional spline characteristics while enabling local shape adjustment through free parameters, with fairing achieved via bending energy minimization. 
Wavelet-based fairing algorithms\cite{LHBAC,RWCMFCSANC} have seen notable development in recent years, attracting extensive research interest globally.
Li et al.\cite{WBMNCF} introduced a multi-resolution fairing framework for NURBS curves using non-uniform semi-orthogonal B-spline wavelets, which provides greater flexibility and applicability than uniform B-spline wavelets and is readily applicable to both 2D and 3D NURBS curves.
Abdul-Rahman et al.\cite{ABDULRAHMAN2013187} developed a lifting wavelet-based fairing algorithm for free-form surfaces, extending its application to regular, semi-regular, and irregular triangular meshes representing non-Euclidean surfaces.
Wang et al.\cite{WBLFASDC} proposed a B-spline surface fairing method that integrates T-splines with wavelet decomposition, offering enhanced local fairing capabilities compared to conventional approaches.
Li et al.\cite{EATSFUBF} proposed an adaptive bilateral filtering method for T-spline fairing that efficiently eliminates serrated artifacts and preserves boundaries by generalizing mesh denoising to continuous surfaces.

The problem is primarily modeled as an energy optimization in existing approaches. However, the computational time and resource requirements of such methods increase significantly when applied to large datasets.
To address this limitation, a curve and surface fairing method based on Progressive-Iterative Approximation (PIA) is proposed. Minor perturbations are systematically applied to enhance geometric fairness, while the introduction of per-control-point fairing weights provides improved manageability and flexibility.

\subsection{Progressive-iterative approximation}

The Progressive-Iterative Approximation (PIA) is a highly effective data fitting method initially introduced and named by Lin et al.\cite{SSJD15103100222743,SJESAFC0B52EF118BE43FA8FEFC84DC20EC7}. By iteratively adjusting control points, PIA generates a sequence of blending curves or surfaces that gradually converge to interpolate or approximate a given set of data points. Inspired by the uniform cubic B-spline gain-loss correction algorithm proposed by Qi et al. \cite{Qi1975,Qi1991} and related work by de Boor\cite{BoorHowDA}, Lin et al. first established the convergence of the non-uniform cubic B-spline iterative algorithm\cite{SSJD15103100222743}, subsequently extending this proof to all positive basis blending curves and surfaces\cite{SJESAFC0B52EF118BE43FA8FEFC84DC20EC7}, thereby formally naming the method Progressive-Iterative Approximation.

Compared to conventional data fitting methods, the Progressive-Iterative Approximation (PIA) algorithm offers greater simplicity, clearer geometric interpretation, and eliminates the need to solve large systems of linear equations\cite{DENG2016403}. Consequently, it has become a significant fitting technique in geometric design\cite{LIN201840}.
In recent years, the PIA framework has been extended in several directions. A weighted PIA variant (WPIA) was introduced by Lu\cite{LU2010129} to accelerate convergence. Lin \cite{LIN2010322} proposed Local PIA (LPIA), which modifies only a subset of control points for adaptive and flexible fitting. Lin et al.\cite{LIN2011967} further developed an Extended PIA (EPIA), enabling parallel computation and allowing fewer control points than data points. Deng and Lin\cite{DENG201432} presented Least-Squares PIA (LSPIA), which performs large-scale fitting efficiently and stably with reduced computational cost. For bicubic B-spline surfaces, Liu et al.\cite{SSJDD85AA5CF334D8425A0DEC8438C6CC006} devised a Jacobi-based PIA algorithm that enhances convergence speed through Jacobi decomposition of the collocation matrix.
Chang et al.\cite{CLSPIAFBCSF} proposed the CLSPIA method, integrating Lagrange multiplier method, Uzawa algorithm, and LSPIA for B-spline fitting with interpolation constraints.

Progressive-Iterative Approximation (PIA) is primarily employed for fitting curves and surfaces to data points. It has also been successfully applied in reverse engineering \cite{IRPIAFCASR,CGPIAFLCSSI}, geometric modeling\cite{APIAMIOA}, data compression \cite{IMFICBULSPIA}, and mesh generation\cite{QGAHMGBACVIFA}. Jiang et al.\cite{JiangFairingPIA2023} introduced a PIA-based fairing method, termed Fairing-PIA, which generates a sequence of faired curves and surfaces that progressively fit the given data points. Furthermore, in recent years, numerous studies have emerged on acceleration algorithms for PIA\cite{OAPAIAMWMFLSF, WLPIAPFTBS, ACLPIAMFCASF}. Research on acceleration algorithms for progressive iterative approximation provides important theoretical support for its industrial applications.

In this paper, the PIA framework is adapted to the fairing of existing curves and surfaces through iterative control point adjustment, while ensuring the resulting geometry remains close to the original shape. Furthermore, it is demonstrated that conventional energy-based fairing models represent a special case of the proposed approach.

\section{Fairing Method}

\subsection{Curve fairing}

In this section, we present the fairing method for B-spline curve. Given a B-spline curve as follows:
\begin{equation}
	C(t) = \sum_{i=1}^{n} N_i(t) P_i, t\in [a, b]
\end{equation}
where $\{P_i\}_{i=1}^n$ are control points, and $ N_i(t)$ is the B-spline basis function  corresponding to the $i$th control point. And the B-spline is defined on the knot vector $\{t_1,t_2,...,t_{n+p+1}\}$, where $p$ is the degree of the B-spline  basis function, and $n$ is the number of the control points. We want to find a fairing curve that approches the original curve $C(t)$(Eq.(1)) as close as possible.

We define the \emph{fairing functional} for the $j$th control point of $C(t)$ (Eq.(1)) as $\mathcal{F}_{j}(f)$, where $f$ is a function, to fair curves by control point adjustment. Its specific formula and derivation process will be provided in section 2.3.

We regard the curve $C(t)$(Eq.(1)) as the initial curve 
\begin{equation}
	C^{[0]}(t) = \sum_{i=1}^{n} N_i(t) P^{[0]}_i, t\in [a, b],
\end{equation}
which means $P^{[0]}_i = P_i,i=1,...,n$.

First, we calculate the \emph{difference vectors} for control points in the first step of the iteration:
\begin{equation}
	\delta_i^{[0]} = P_i-P_i^{[0]}, i=1,...,n.
\end{equation}

Next, the \emph{fairing vectors} for control points of the first iteration can be calculated as
\begin{equation}
	\eta_i^{[0]} = \sum_{l=1}^{n} \mathcal{F}_{i}(N_l(t)) P_l^{[0]}, i=1,...,n.
\end{equation}    

Then, the new control points $\{P_i^{[1]}\}_{i=1}^n$ after the first iteration comprise the contribution of two terms:
\begin{equation}
	P^{[1]}_i = P_i^{[0]} +\mu_i[(1-\omega_i)\delta^{[0]}_i-\omega_i\eta^{[0]}_i],i=1,...,n,
\end{equation} 
where $\mu_i$ is a normalization weight, and $\omega_i$ is a fairing weight corresponding to the $i$th control points. The larger the fairing weight $\omega_i$ is, the fairer the curve is. But at the same time, the deviation of the control points is also larger.

The new curve is obtained as follows by substituting the new control points into curve(Eq.(1)):
\begin{equation}C^{[1]}(t) = \sum_{i=1}^{n} N_i(t) P^{[1]}_i , t\in [a, b].\end{equation} 

This procedure is performed iteratively. After the $k$th iteration, the $k$th curve $P^{[k]}(t)$ is generated:
\begin{equation}C^{[k]}(t) = \sum_{i=1}^{n} N_i(t) P^{[k]}_i, t\in [a, b].\end{equation}

We first calculate the $(k+1)$th difference vectors for control points,
\begin{equation}\delta_i^{[k]} = P_i - P_i^{[k]}, i=1,...,n. \end{equation}

Next, the fairing vectors for control points are calculated by
\begin{equation}\eta_i^{[k]} = \sum_{l=1}^{n}\mathcal{F}_{i}(N_l(t)) P_l^{[k]},i=1,...,n.\end{equation}

Then the new control points in the $(k+1)$th iterative are
\begin{equation}
	P^{[k+1]}_i = P_i^{[k]} +\mu_i[(1-\omega_i)\delta^{[k]}_i-\omega_i\eta^{[k]}_i],i=1,...,n.
\end{equation} 

Finally, the $(k+1)$th curve is 
\begin{equation}C^{[k+1]}(t) = \sum_{i=1}^{n} N_i(t) P^{[k+1]}_i , t\in [a, b].\end{equation}

In this way, we obtain a sequence of curves $\{C^{[k]}(t), k=1,2,3,...\}$. In the next section, we will prove that the sequence of curves converges. 

During the iteration of the fairing algorithm, if we do not restrict the location of control points,  the curve always ends up tending to a straight line. The deviation value of control points is taken into account when the difference vector of control points is set which ensures that the deviation value of control points is controllable in the process of curve fairing. In the iterative format Eq.(10), $\mu_i$ is set to ensure the convergence of the algorithm, and $\omega_i$ is set to balance the the deviation of the control point with the fairing vector. In this way, our algorithm is more flexible and geometrically meaningful than the traditional energy minimization algorithm.

\subsection{Surface fairing}
The fairing method for curve fairing can be easily extend to the surface case. In this section, we present the details of our method for surface fairing.

Given a B-spline surface as follows:
\begin{equation}
		S(u,v) = \sum_{i=1}^{n_1}\sum_{j=1}^{n_2} N_i(u) N_j(v) P_{ij}, 
		(u,v)\in [a_1, b_1]\times[a_2,b_2] 
\end{equation}
where $\{P_{ij}^{[0]}\}_{i=1,j=1}^{n_1,n_2}$ are control points, and $ N_i(u)$ and $N_j(v)$ are the B-spline basis functions. 

We rearrange the control points and their B-spline basis functions in lexicographical order as follows:
$$[P_1,P_2,...,P_{n_1 n_2}] = [P_{11}, P_{12}, ...,P_{n_1,n_2}], $$
$$[N_1(u,v), N_2(u,v),...,N_{n_1n_2}(u,v)]=[N_1(u)N_1(v), N_1(u)N_2(v),...,N_{n_1}(u)N_{n_2}(v)]. $$

We define the \emph{fairing functional} for the $j$th control point as $\mathcal{F}_{j}(f)$, where $f$ is a function, to fair surfaces by control point adjustment. Their specific formulas and derivation process will also be provided in section 3.3.

Then, the initial surface can be rewritten as
\begin{equation}
		S^{[0]}(u,v) = \sum_{i=1}^{n_1 n_2} N_i(u,v) P_i^{[0]},
		(u,v) \in [a_1, b_1]\times[a_2,b_2] 
\end{equation}

The $k$th surface $S^{[k]}(u,v)$ can be generated after the $k$th iteration similar to the curve case. In this iteration, we first calculate the $(k+1)$th diffrenece vectors for control points:
\begin{equation}\delta_i^{[k]} = P_i - P_i^{[k]}, i=1,2,...,n_1 n_2.\end{equation}
Then , the fairing vectors for control points are calculated by
\begin{equation}\eta_i = \sum_{l=1}^{n_1 n_2}\mathcal{F}_{i}(N_l(u,v)) P_l^{[k]}, i=1,2,...,n_1 n_2.\end{equation}
Finally, the new control points of the $(k+1)$th surface are produced by
\begin{equation}
	P^{[k+1]}_i = P_i^{[k]} +\mu_i[(1-\omega_i)\delta_i^{[k]}-\omega_i\eta_i^{[k]}], i=1,2,...,n_1 n_2
\end{equation}
which leads to the $(k+1)$th B-spline surface:
\begin{equation}
	S^{[k+1]}(u,v) = \sum_{i=1}^{n_1n_2}N_i(u,v)P^{[k+1]}_i.
\end{equation}

The parammeters $\omega_i$ and $\mu_i$ here are similar to those in section 2.1.

In this way, we obtain a sequence of surfaces $\{S^{[k]}(u,v),k=1,2,3,...\}$. The convergence analysis of the surface fairing for surface case is analogous to that of the curve case in section 3.1.

\subsection{Fairing functional}

In engineering practice, designers evaluate curve fairness by observing the curvature distribution. If the curvature of the curve changes uniformly, the curve is considered fair\cite{Burchard19941AW}. Since fairness involves geometric appearance and is greatly affected by subjective factors, it is difficult to quantify curve fairness with a unified quantitative index. 

We usually use the following energy model to approximate fairing model: 
\begin{equation}
	E(C) =\int_{a}^{b}||C^{(r)}(t) ||^2dt, 
\end{equation}
where $r=1,2,3$. $C^{(r)}(t)$ is the $r$th derivative of the curve $C(t)$. 
Eq.(18) corresponds to three types of energy: stretch energy\cite{Veltkamp1995Modeling3C}, strain energy\cite{ZHANG2001913}, and jerk energy\cite{MEIER1987297} when $r=1,2,3$.

With respect to the three types of energies(Eq.(18)), we define the fairing functional as
\begin{equation}
	\mathcal{F}_{j}(f) = \int_{a}^{b} N'_j(t) f' dt, \ j=1,...,n,
\end{equation}
or
\begin{equation}
	\mathcal{F}_{j}(f) = \int_{a}^{b} N''_j(t) f'' dt, \ j=1,...,n,
\end{equation}
or
\begin{equation}
	\mathcal{F}_{j}(f) = \int_{a}^{b} N^{(3)}_j(t) f^{(3)} dt, \ j=1,...,n.
\end{equation}

Similar to the curve case, we usually use the following energy functional to approximately evaluate the fairness of a surface $S(u,v)$\cite{WANG1997485}:
\begin{equation} 
	E(S) = \int_{a_1}^{b_1}\int_{a_2}^{b_2}||S_{u}(u,v)||^2  + ||S_{v}(u,v)||^2 dudv,
\end{equation}
or
\begin{equation}
	\begin{aligned}
		E(S) &= \int_{a_1}^{b_1}\int_{a_2}^{b_2}||S_{uu}(u,v)||^2  + 2||S_{uv}(u,v)||^2 + ||S_{vv}(u,v)||^2dudv. 
	\end{aligned}
\end{equation}

We define the following fairing functionals analogous to the curve case corresponding to Eq.(22):
\begin{equation}
	\begin{aligned}
		\mathcal{F}_{j}(f) = \mathcal{F}_{u, j}(f)+\mathcal{F}_{v, j}(f),\\
		\mathcal{F}_{u, j}(f) = \int_{a_1}^{b_1}\int_{a_2}^{b_2} N_{u,j}(u,v) f_u dudv,\\
		\mathcal{F}_{v, j}(f) = \int_{a_1}^{b_1}\int_{a_2}^{b_2} N_{v,j}(u,v) f_v dudv
	\end{aligned}
\end{equation}
where $N_{u,j}(u,v),  N_{v,j}(u,v)$ represent the first-order partial derivatives of $N_j(u,v)$ with respect to $u$ and $v$. 

And we define the following fairing functionals corresponding to Eq.(23):
\begin{equation}
	\begin{aligned}
		\mathcal{F}_{j}(f) = \mathcal{F}_{uu, j}(f)+2\mathcal{F}_{uv, j}(f) + \mathcal{F}_{vv, j}(f),\\
		\mathcal{F}_{uu, j}(f) = \int_{a_1}^{b_1}\int_{a_2}^{b_2} N^{(u,u)}_j(u,v) f^{(u,u)} dudv,\\
		\mathcal{F}_{uv, j}(f) = \int_{a_1}^{b_1}\int_{a_2}^{b_2} N^{(u,v)}_j(u,v) f^{(u,v)} dudv,\\
		\mathcal{F}_{vv, j}(f) = \int_{a_1}^{b_1}\int_{a_2}^{b_2} N^{(v,v)}_j(u,v) f^{(v,v)} dudv
	\end{aligned}
\end{equation}
where $N^{(u,u)}_j(u,v)$,  $N^{(u,v)}_j(u,v)$ and $N^{(v,v)}_j(u,v)$ represent the second-order partial derivatives of $N_j(u,v)$ with respect to $u$ and $v$. 

The fairing functionals corresponding to other energy functionals can be definded similarly.

\subsection{Automatic Selection of Control Points}

Our algorithm supports the fairing of user-specified portions and, simultaneously, provides optimal control point modification options to accommodate automated fairing schemes. We first rank each control point based on the degree of energy difference function of curves and surfaces before and after fairing. Subsequently, the top-ranked control points are adjusted using the methods mentioned in Sections 2.1 and 2.2 to obtain fairing curves and surfaces. A brief description is as follows:

In the case of curves, if a single control point $P_j$ is adjusted to achieve curve fairing, we can let $\frac{\partial E(C)}{\partial P_j} = 0$, which means
\begin{equation}
	\hat{P}_j = \frac{-\sum_{i\neq j}\mathcal{F}_{j}(N_i(t)) P_i}{\mathcal{F}_{j}(N_j(t))}.
\end{equation}

Literature \cite{1995Localcurve} presented a method for sorting curve control points. Define  $Z_j$ to measure the impact of adjusting control point  $P_j$ on curve fairing:
\begin{equation}
	Z_j = E(C) - E(\hat{C}),
\end{equation}
where $\hat{C}(t) =  \sum_{i=1,i\neq j}^{n} N_i(t) P_i + N_j(t) \hat{P}_j, t\in [a, b]$.

Unlike the method in \cite{1995Localcurve}, here we can select multiple control points as active control points to perform fairing-based progressive iterative approximation.

$E(C)$(Eq.(18)) can be derived into the following form:

\begin{equation}
	\begin{aligned}
		E(C)&= \int_{a}^{b} ||C^{(r)}(t) ||^2 dt\\
		&=\int_{a}^{b} ||\sum_{i=1}^n N_i^{(r)}(t)P_i ||^2 dt\\
		&=\int_{a}^{b} (\sum_{i=1}^n N_i^{(r)}(t)P_i)^T(\sum_{i=1}^n N_i^{(r)}(t)P_i) dt\\
		&=\int_{a}^{b} (\sum_{i\neq j}^n N_i^{(r)}(t)P_i)^T(\sum_{i\neq j}^n N_i^{(r)}(t)P_i) + 2N_j^{(r)}(t)P_j^T\sum_{i\neq j}^n N_i^{(r)}(t)P_i +(N_j^{(r)}(t))^2P_j^TP_j dt
	\end{aligned}
\end{equation} 

\begin{algorithm}[H]
	\caption{ Curve Fairing Method with Automatic Selection of Control Points }\label{algorithm1}
	\begin{algorithmic}[1]
		\Require{initial curve $C(t)$, the number of adjusted control points $m$, fairing weights$\{\omega_{i}\}_{i=1}^{m}$, max iterative time $K$, upper bound $\epsilon$ for the absolute difference in control point deviations between two iterations}
		\Ensure{fairing curve $\hat{C}(t)$}
		\State $\bar{Z} = [0, 0, \ldots, 0], idx = [1, 2, \ldots, n]$\;
		\For {$j = 1$ to $n$}
		\State {$Z_j = \frac{||\mathcal{F}_j(C(t))||^2}{\mathcal{F}_j(N_j(t))} ;$}
		\EndFor
		\State	Sort $idx$ in descending order based on corresponding values in $\bar{Z}$ {Maintain index-value correspondence}\;
		\State$I = \{ idx[1], idx[2], \ldots, idx[m] \}$ {Select first $m$ indices}\;
		\State$k\leftarrow 0$\;
		\While{$k<K$}
		\State $P^{[k+1]}_i = P_i^{[k]} +\mu_i[(1-\omega_i)\delta^{[k]}_i-\omega_i\eta^{[k]}_i],i\in I$\;
		\State$	P^{[k+1]}_i = P_i, i \notin I$\;
		\State$C^{[k+1]}(t) = \sum_{i=1}^{n} N_i(t) P^{[k+1]}_i , t\in [a, b]$\;
		\State$E^{[k+1]}_{iter} =  \sqrt{\frac{\sum_{i\in I}||P_i^{[k+1]}-P_i^{[k]}||^2}{\sum_{i\in I}||P_i^{[k+1]}-P_i^{[0]}||^2}}$\;
		\If{$|E^{[k+1]}_{iter} - E^{[k]}_{iter}| < \Delta p$}
		\State break\;
		\EndIf 
		\State $k\leftarrow k+1$\;
		\EndWhile
	\end{algorithmic}
\end{algorithm}

According to Eq.(26), we have
\begin{equation}
		P_j - \hat{P}_j =  P_j + \frac{\sum_{i\neq j}\mathcal{F}_{j}(N_i(t)) P_i}{\mathcal{F}_{j}(N_j(t))}=\frac{\sum_{i=1}^{n}\mathcal{F}_j(N_i(t))P_i}{\mathcal{F}_j(N_j(t))}
\end{equation} 

Then $Z_j$ (Eq.(27)) can be derived into the following form

\begin{equation}
	\begin{aligned}
		&Z_j  \\
		&= E(C) - E(\hat{C}) \\
		&=\int_{a}^{b} 2N_j^{(r)}(t)(P_j-\hat{P}_j)^T\sum_{i\neq j}^n N_i^{(r)}(t)P_i +(N_j^{(r)}(t))^2(P_j^TP_j-\hat{P}_j^T\hat{P}_j) dt\\
		&=\int_{a}^{b} N_j^{(r)}(t)(P_j-\hat{P}_j)^T\left[2\sum_{i\neq j}^n N_i^{(r)}(t)P_i + N_j^{(r)}(t)(P_j+\hat{P}_j)\right]dt\\
		&= \int_{a}^{b} N_j^{(r)}(t)(P_j-\hat{P}_j)^T\left[2\sum_{i=1}^n N_i^{(r)}(t)P_i + N_j^{(r)}(t)(\hat{P}_j - P_j)\right]dt\\
		&=(P_j-\hat{P}_j)^T \left[\sum_{i=1}^n 2\int_{a}^{b} N_j^{(r)}(t)N_i^{(r)}(t)P_i dt + 
		 \int_{a}^{b}  (N_j^{(r)}(t))^2(t)dt(\hat{P}_j - P_j)\right] \\
		&=(P_j-\hat{P}_j)^T \left[\sum_{i=1}^n 2\mathcal{F}_j(N_i(t))P_i - \mathcal{F}_j(N_j(t))(P_j-\hat{P}_j)^T\right]\\
		&= \frac{\sum_{i=1}^{n}\mathcal{F}_j(N_i(t))P_i^T}{\mathcal{F}_j(N_j(t))} \sum_{i=1}^n \mathcal{F}_j(N_i(t))P_i \\
		&=\frac{||\sum_{i=1}^{n}\mathcal{F}_j(N_i(t))P_i||^2}{\mathcal{F}_j(N_j(t))}\\
		&=\frac{||\mathcal{F}_j(C(t))||^2}{\mathcal{F}_j(N_j(t))}
	\end{aligned}
\end{equation}

The value of $Z_j$ quantifies the magnitude of change in the curve's fairing energy induced by modifying this specific control point. A larger $Z_j$ value indicates a higher importance of the point in developing the fairness of the curve. 


Given an initial B-spline curve $C(t)$ as Eq.(1) and the number of adjusted control points $m$. The specific procedure is as follows:
\begin{enumerate}
	\item Energy Difference Calculation: Calculate $Z_i$ corresponding to each control point $P_i$ according to Eq.(30). 
	\item Ranking by Energy Difference: After calculating the energy differences $\bar{Z}_i$ for all points, all control points are sorted in descending order based on their $\bar{Z}_i$ values. Consequently, the points at the top of the sorted sequence are the control points with the largest energy differences, representing those most in need of adjustment.
	\item Active Control Points Selection: The top $m$ control points with the largest energy differences are selected from the sorted sequence as the active control points. The set of indices for these selected points is denoted as $I$.
	\item Iterative Adjustment: This process iteratively updates control points, analogous to the algorithm in Section 3.1, with the distinction that updates are applied exclusively to the active control points. In the $k$th iteration, calculate the difference vectors of active points
	$$\delta_i^{[k]} = P_i - P_i^{[k]}, i\in I$$
	and the fairing vectors of active points
	$$\eta_i^{[k]} = \sum_{l=1}^{n}\mathcal{F}_{i}(N_l(t)) P_l^{[k]}, i\in I.$$
	Then the new control points in the (k+1)th iterative are
	$$	P^{[k+1]}_i = P_i^{[k]} +\mu_i[(1-\omega_i)\delta^{[k]}_i-\omega_i\eta^{[k]}_i], i\in I,$$
	$$ P^{[k+1]}_i = P_i, i\notin I.$$
	The $(k+1)th$ curve is 
	$$
		C^{[k+1]}(t) = \sum_{i=1}^{n} N_i(t) P^{[k+1]}_i = \sum_{i\in I}^{n} N_i(t) P^{[k+1]}_i + \sum_{i\notin I}^{n} N_i(t) P_i.
	$$
\end{enumerate}

The pseudocode is detailed in Algorithm \ref{algorithm1}.
\\

In the case of surfaces, if a single control point $P_j$ is adjusted to achieve surface fairing, we can let $\frac{\partial E(C)}{\partial P_j} = 0$, which means
$$\hat{P}_j = \frac{-\sum_{i\neq j}\mathcal{F}_{j}(N_i(u,v)) P_i}{\mathcal{F}_{j}(N_j(u,v))} .$$

We define $Z_j$ to measure the impact of adjusting control point $P_j$  on surface fairing:
\begin{equation}
	Z_j = E(S) - E(\hat{S}),
\end{equation}
where $$\hat{S}(u,v)=\sum_{i\neq j}N_i(u,v)P_i+N_j(u,v)\hat{P}_j, t\in [a, b].$$

Similarly, it can be easily derived that
\begin{equation}
	Z_j =  \frac{||\mathcal{F}_{j}(S(u,v))||^2}{\mathcal{F}_{j}(N_j(u,v))}.
\end{equation}

Here $Z_j$ measures the impact of adjusting control point $P_j$ on surface fairing.  we need to calculate the value of $Z_j$ and select th top $m$ values in descending order as the active points of the surface.


The surface algorithm is similar to the curve algorithm, so it will not be detailed here.

\section{Numerical analysis}

\subsection{Convergence analysis}

In this section, we show that the  curve fairing method is convergent.

First, we rewritten the iterative format (Eq.(10)) into the following matrix form:
\begin{equation}
	\begin{aligned}
		P^{[k+1]} &= P^{[k]} +  \Lambda[(I-\Omega) (P - P^{[k]}) -\Omega D_r  P^{[k]}]\\
		&= [ I- \Lambda (I - \Omega + \Omega D_r) ] P^{[k]} +\Lambda (I - \Omega) P\\
		&= [ I- \Lambda A ] P^{[k]} +\Lambda (I - \Omega)P
	\end{aligned}
\end{equation} 
in which 
\begin{equation*}
	\begin{aligned}
		P^{[k]} &=\begin{bmatrix}	P_{1}^{[k]}  & P_{2}^{[k]}& \cdots  & P_{n}^{[k]} \end{bmatrix}^T, \\
		P &= \begin{bmatrix}	 P_{1} & P_{2}& \cdots  & P_{n} \end{bmatrix}^T,\\
		\Omega &= diag(\omega_1, \omega_2 ,...,\omega_n), \\
		\Lambda &= diag(\mu_1, \mu_2, ...,\mu_n),
	\end{aligned}
\end{equation*}

\begin{equation} A = I - \Omega + \Omega D_r, \end{equation}
and
\begin{equation}
	D_r=\begin{bmatrix}
		\mathcal{F}_{1}(N_1(t))& \cdots  & 	\mathcal{F}_{1}(N_n(t)) \\
		\mathcal{F}_{2}(N_1(t)) & \cdots  & 	\mathcal{F}_{2}(N_n(t))\\
		\vdots                              & \ddots       & \vdots\\
		\mathcal{F}_{n}(N_1(t)) & \cdots  & \mathcal{F}_{n}(N_n(t))
	\end{bmatrix}.
\end{equation} 

Moreover, $diag(\cdot)$ denotes the diagonal matrix, and $r$ can be chosen from the set $\{1,2,3\}$.

{\bf Lemma 4.1} {\it Let $A=[a_{ij}]$(Eq.(34)) be a strict diagonally dominant matrix. If the diagonal elements of matrix $\Lambda$ satisfy $0<\mu_i <\frac{2}{\sum_{j=1}^{n}|a_{ij}|}, i=1,...,n$, and $\omega_i$ statifies $0<\omega_i <1$, then we obtain $0<||I-\Lambda A||_{\infty} <1$.} 

\begin{proof}
	Based on the definition of strictly diagonally dominant matrix, we obtain $|a_{ii}| > \sum_{j\neq i}|a_{ij}|\geq 0$. Since $ a_{ii}  = (1-\omega_i) + \omega_i  \int_{a}^{b} (N^{(r)}_i(t))^2 dt > 0$, we have\\
	(1) If $0<\mu_i \leq\frac{1}{a_{ii}}$, then we have
	$$ ||I-\Lambda A||_{\infty} = \max_{i}(1-\mu_i a_{ii}+\mu_i\sum_{j\neq i}|a_{ij}|) $$
	$$ <\max_i (1-\mu_i a_{ii} + \mu_i a_{ii}) =  1$$
	(2) If $\frac{1}{a_{ii}} <\mu_i<\frac{2}{\sum_{j=1}^n|a_{ij}|}$, then we have
	$$ ||I-\Lambda A||_{\infty} = \max_{i}(\mu_i a_{ii} - 1 +\mu_i\sum_{j\neq i}|a_{ij}|) $$
	$$ =\max_i (\mu_i\sum_{j=1}^{n}|a_{ij}|-1)$$
	$$ < \max_{i}(\frac{2}{\sum_{j=1}^{n}|a_{ij}|}\sum_{j=1}^{n}|a_{ij}|-1) =1 $$
	
	Consequently, when  $0<\mu_i <\frac{2}{\sum_{j=1}^{n}|a_{ij}|}, i=1,...,n$, we obtain $0<||I-\Lambda A||_{\infty} <1$ by combining the results of case (1)(2).
\end{proof}

{\bf Lemma 4.2} {\it If $\omega_i,i=1,...,n$ statifies $0<\omega_i <\min\{\frac{1}{2},\frac{1}{4p\max_{j}|d_{ij}|}\}$, $d_{ij}$ is the element of matrix $D_r=[d_{ij}]$(Eq.(35)), $p$ is the degree of the curve, then $A=[a_{ij}]$(Eq.(34)) is a strict diagonally dominant matrix.}

\begin{proof} Since $0<\omega_i <\min\{\frac{1}{2},\frac{1}{4p\max_{j}|d_{ij}|}\}$, and $d_{ii} = \int_{a}^{b} (N^{(r)}_i(t))^2 dt > 0$, 
	we obtain that 
	$$a_{ii}  = (1-\omega_i) + \omega_i d_{ii} > 1-\omega_i >\frac{1}{2}$$
	Besides, we have
	$$|\omega_i d_{ij}| =  \omega_i |d_{ij}|< \omega_i \max_{j} |d_{ij}| <\frac{1}{4p}. $$
	
	According to the local support property of B-spline basis functions, $N_i(t) = 0$ when $t\notin [t_i,t_{i+p+1})$. Obviously $N^{(r)}_i(t) = 0, r\leq p$ when $t\notin [t_i,t_{i+p+1})$. It can be inferred that  when $|i-j|\geq p+1 $, $\forall t\in [t_1,t_{n+p+1}]$, $N^{(r)}_i(t) N^{(r)}_j(t) =0$ which also means $d_{ij} = 0$. We can easily derive that the matrix has at most 
	$2p+1$ non-zero values in each row. 
	
	Hence, it can be obtained that 
	$$\sum_{j\neq i}|a_{ij}| = \sum_{j\neq i}|\omega_i d_{ij}| <\frac{(2p+1)-1}{4p}  = \frac{1}{2}<a_{ii}.$$
	
	Consequently, when $\omega_i,i=1,...,n$ statifies $0<\omega_i <\min\{\frac{1}{2},\frac{1}{4p\max_{j}|d_{ij}|}\}$, $A=[a_{ij}]$(Eq.(34)) is a strict diagonally dominant matrix.
\end{proof}

{\bf Theorem 4.3} {\it When we choose $\omega_i,i=1,...,n$ statifies $0<\omega_i <\min\{\frac{1}{2},\frac{1}{4p\max_{j}|d_{ij}|}\}$, the curve fairing method(26) is convergent.}

\begin{proof} We obtain the iterative method from Eq.(10) as follows:
	\begin{equation}
		\begin{aligned}
			P^{[k+1]} =& P^{[k]} +  \Lambda[(I-\Omega) (P - P^{[k]}) -\Omega D_r  P^{[k]}]\\
			=& (I- \Lambda A)P^{[k]} +\Lambda (I - \Omega) P\\
			=& (I- \Lambda A)^2P^{[k-1]}  +\Lambda (I - \Omega) P \\
			&+ ( I- \Lambda A)\Lambda (I - \Omega) P\\
			=& \cdots \\
			=&(I- \Lambda A)^{k+1}P^{[0]}  + \sum_{l=0}^{k}(I- \Lambda A)^l\Lambda (I - \Omega) P\\
			=&(I- \Lambda A)^{k+1}P  + \sum_{l=0}^{k}(I- \Lambda A)^l\Lambda (I - \Omega) P
		\end{aligned}
	\end{equation} 
	
	Based on Lemma 4.1 and Lemma 4.2, the spectral radius of $I- \Lambda A$ satisfies 
	$$ 0<\rho(I-\Lambda A) <||I-\Lambda A || <1, $$
	and we obtain 
	$$\lim_{k\to \infty} ( I- \Lambda A)^{[k+1]} = O, \lim_{k\to \infty} \sum_{l=0}^{k}(I- \Lambda A )^l = \Lambda (\Lambda A)^{-1}.$$
	
	Hence, when $k\to \infty$, Eq.(37) tends to
	$$P^{[\infty]} = (\Lambda A)^{-1}\Lambda (I-\Omega)P  = A^{-1}(I-\Omega)P$$
	which is the solution of the linear system as follows,
	\begin{equation}A P^{[\infty]} = (I-\Omega)P.\end{equation}
\end{proof}

{\bf Remark 4.4} {\it We select the diagonal elements of matrix $\Lambda$ as $\mu_i = \frac{1}{\sum_{j=1}^{n}|a_{ij}|},  i=1,2,...,n$ in pratice.}

\subsection{Relation to traditional model}
In this section, we will show that the traditional energy method is a special case of our fairing method.

In a special case of $\Omega = \omega I$, i.e., all fairing weights $\omega_i$, $j=1,2,...,n$(Eq.(5)) are equal to $\omega$, our method(Eq.(10)) degenerates to 
$$P^{[k+1]} = P^{[k]} +  \Lambda[(1-\omega) (P - P^{[k]}) -\omega D_r  P^{[k]}]$$
$$= [ I- \Lambda B ] P^{[k]} +\Lambda (1 - \omega) P$$
where
\begin{equation}B = (1 - \omega)I + \omega D_r \end{equation}

In the traditional energy minimization fairing method, the goal is to find the minimum of the energy functional, which is defined as follows:
\begin{equation}E = \frac{1-\omega}{2} f_1 + \frac{\omega}{2} f_2 \end{equation}
where $f_1$ and $f_2$ represent the deviation term and the fairing term, respectively, defined as
\begin{equation}f_1 = \sum_{i=1}^n || \hat{P}_i - P_i||^2 \end{equation}
\begin{equation}f_2 = \int || \hat{C}^{(r)}(t)||^2 dt\end{equation}
where $\hat{C}^{(r)}(t)$ represents the $r$th derivative of the fairing curve and $\hat{P}_i$ represents the $i$th control points. $\omega$ is the fairing weight. The larger the fairing weight $\omega$ is, the fairer the generated curve is.

The optimization problem becomes
$$  \min_{\hat{P}_j} E = \min_{\hat{P}_j} \left[\frac{1-\omega}{2} \sum_{i=1}^n || \hat{P}_i - P_i||^2 + \frac{\omega}{2}  \int || \hat{C}^{(r)}(t)||^2 dt \right]  $$
when substituting Eq.(40) and Eq.(41) into Eq.(39).

We obtain the following by $\frac{\partial E}{\partial \hat{P}_j} = 0$, with $j=1,2,...,n$:
$$\frac{\partial E}{\partial \hat{P}_j} =(1-\omega)(\hat{P}_j - P_j) +$$
$$ \omega \sum_{i=1}^{n} \int   N_i^{(r)}(t) N_{j}^{(r)}(t) dt \ \hat{P}_i = 0.$$
Then we have
$$(1-\omega)\hat{P}_j + \omega \sum_{i=1}^{n} \int   N_i^{(r)}(t) N_{j}^{(r)}(t) dt \ \hat{P}_i  = (1-\omega) P_j$$
which can be represented by a matrix as
\begin{equation}
	\begin{aligned}
		B\hat{P} = (1-\omega)P 
	\end{aligned}
\end{equation}
where $B = (1 - \omega)I + \omega D_r $(see Eq.(38)). A fairing curve that minimizes energy model(32) can be obtained after the equation is solved and the obtained control points are substituted into the curve equation.

\section{Experiments and discussions}

In this section, some numerical examples are presented to demonstrate the effictiveness of our algorithm. At the same time, it will be compared with the traditional method of establishing energy equation to show the advantages of our algorithm. In the numerical examples, we employ the root mean squared distance errors as the deviation value of the control points
\begin{equation}
	E^{[k]}_{dev} = \sqrt{\frac{\sum_{i=1}^{n}||P_i^{[k]}-P_i||^2}{n}}
\end{equation}
and
\begin{equation}
	E_{eng,abs}^{[k]} = \int_{a}^{b}||(C^{[k]}(t))^{(r)}||^2dt, r=1,2,3,
\end{equation}
as the absolute energy to measure the deviation and fairness of the curve.

Moreover, we define the relative iteration deviation value as
\begin{equation}
	E^{[k]}_{iter} =  \sqrt{\frac{\sum_{i=1}^{n}||P_i^{[k]}-P_i^{[k-1]}||^2}{\sum_{i=1}^{n}||P_i^{[k]}-P_i^{[0]}||^2}},
\end{equation}
and define the relative energy as
\begin{equation}
	E^{[k]}_{energy,rel} =  \frac{E_{eng,abs}^{[k]}}{E_{eng,abs}^{[0]}}.
\end{equation}

In our implementation, the iteration stops when $| E^{[k+1]}_{iter}- E^{[k]}_{iter}|<10^{-6}$ or when $k>800$.

In the case of a surface, the measurement of fairness and the iterative stop condition are similar to those of a curve.

We implemented our algorithm in C++ on a PC with an Intel Core i9-12900H 2.50GHz CPU and 16 GB RAM.

\subsection{Curve}

This section demonstrates the effectiveness of the proposed algorithm by applying it to three models: a Dolphin, Shape G, and an Archimedean spiral, with comparisons made against the traditional energy method. The surface fairing effects and convergence speeds under different fairing functionals are also evaluated and compared.

\begin{figure}[H]
\centering
\subfigure[]{
	\begin{minipage}[b]{.3\linewidth}
		\centering
		\includegraphics[scale=0.25]{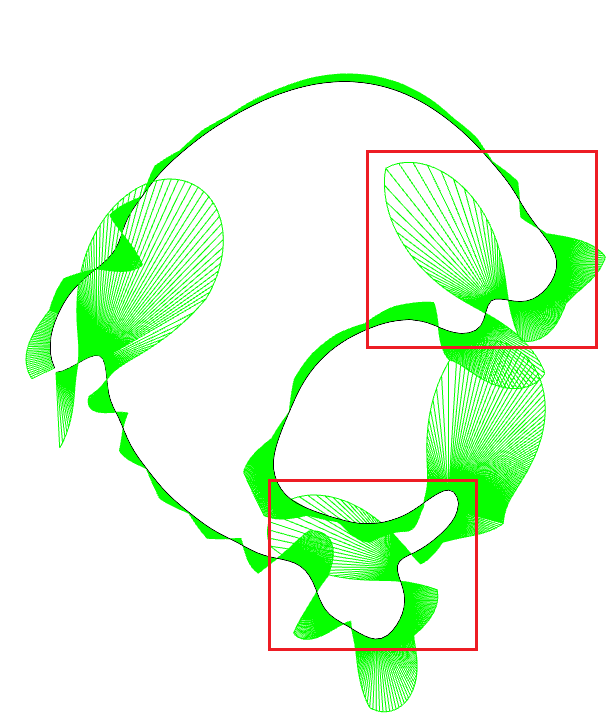}
	\end{minipage}
}
\subfigure[]{
	\begin{minipage}[b]{.3\linewidth}
		\centering
		\includegraphics[scale=0.25]{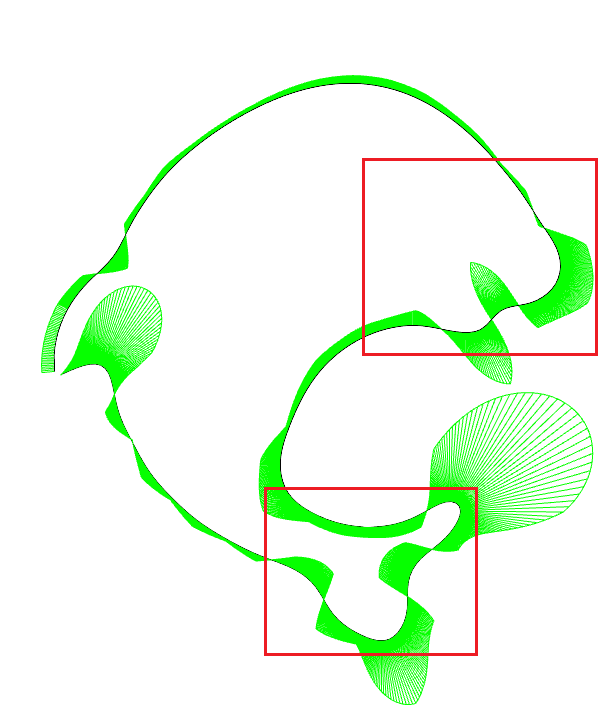}
	\end{minipage}
}
\subfigure[]{
	\begin{minipage}[b]{.3\linewidth}
		\centering
		\includegraphics[scale=0.25]{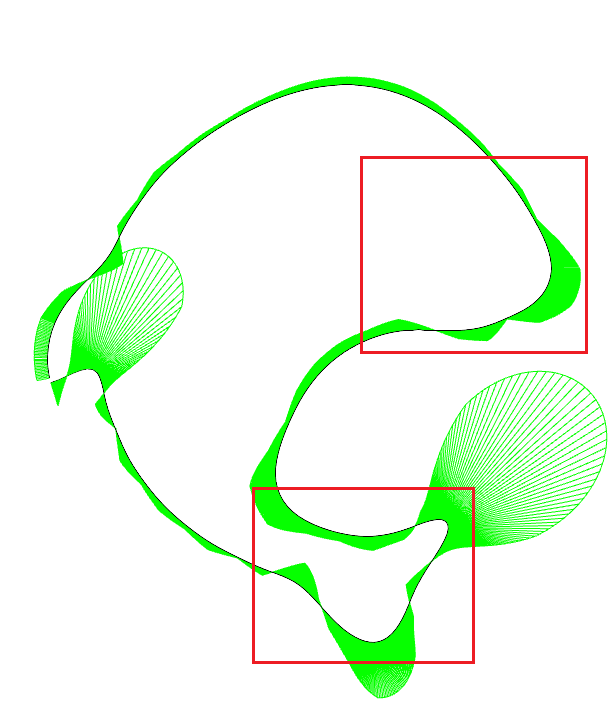}
	\end{minipage}
}

\caption{\label{fig1}Comparison results for the Dolphin model. We add the  curvature comb to show the fairness of the curves. (a) Original curve. (b) The curve fairing by the energy method. (c) The curve fairing by our method. }
\end{figure}

In the example of the Dolphin model, the original curve (Figs.\ref{fig1}(a)) resembles a dolphin, and according to its curvature comb, it can be seen that the fairness of the curve needs to be improved. We set $\omega = 1\times 10^{-6} $ and $r=2$ in energy functional (Eq.(39)) to obtain a fairing curve(Figs.\ref{fig1}(b)) by the traditional energy mothod. Overall, the fairness of the curve has been improved, but it is clear that the fairness of the head and tail of the "dolphin" (in the red box) still does not meet the requirement. 

We set the fairing weights of the 25th to 32th control points as $ 1\times 10^{-5} $ and the 46th to 57th control points as $8\times10^{-6}$ , thereby affecting the curve segment with undesirable fairness. The other control points' fairing weights are still set as $1\times 10^{-6}$, and the fairing curve is generated by our method(Figs.\ref{fig1}(c)). Evidently, our method allows for more precise adjustments to the shape of the current curve, resulting in a model with better fairness and refinement.

\begin{figure}[H]
\centering
\subfigure[]{
	\begin{minipage}[b]{.3\linewidth}
		\centering
		\includegraphics[scale=0.22]{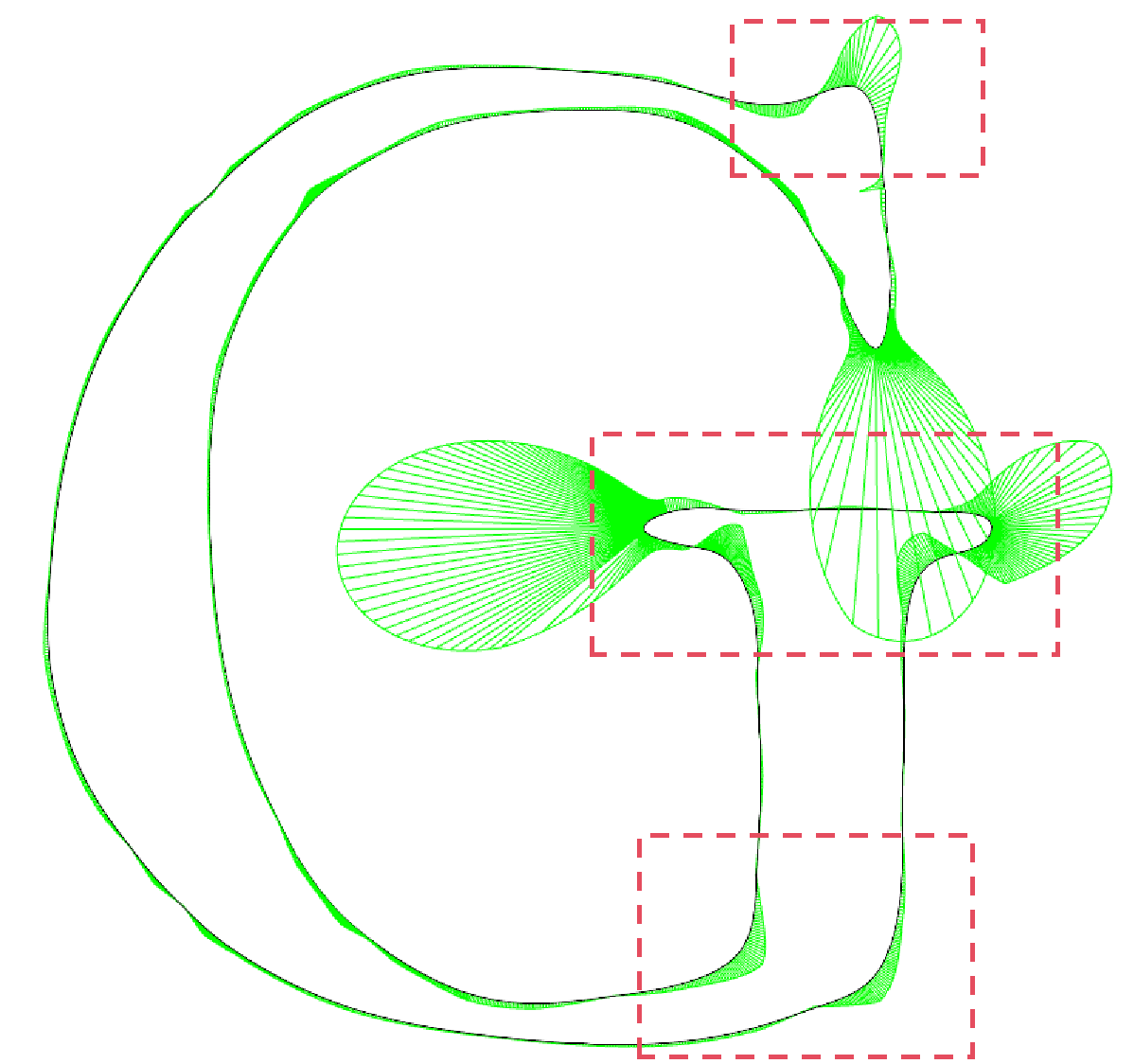}
	\end{minipage}
}
\subfigure[]{
	\begin{minipage}[b]{.3\linewidth}
		\centering
		\includegraphics[scale=0.22]{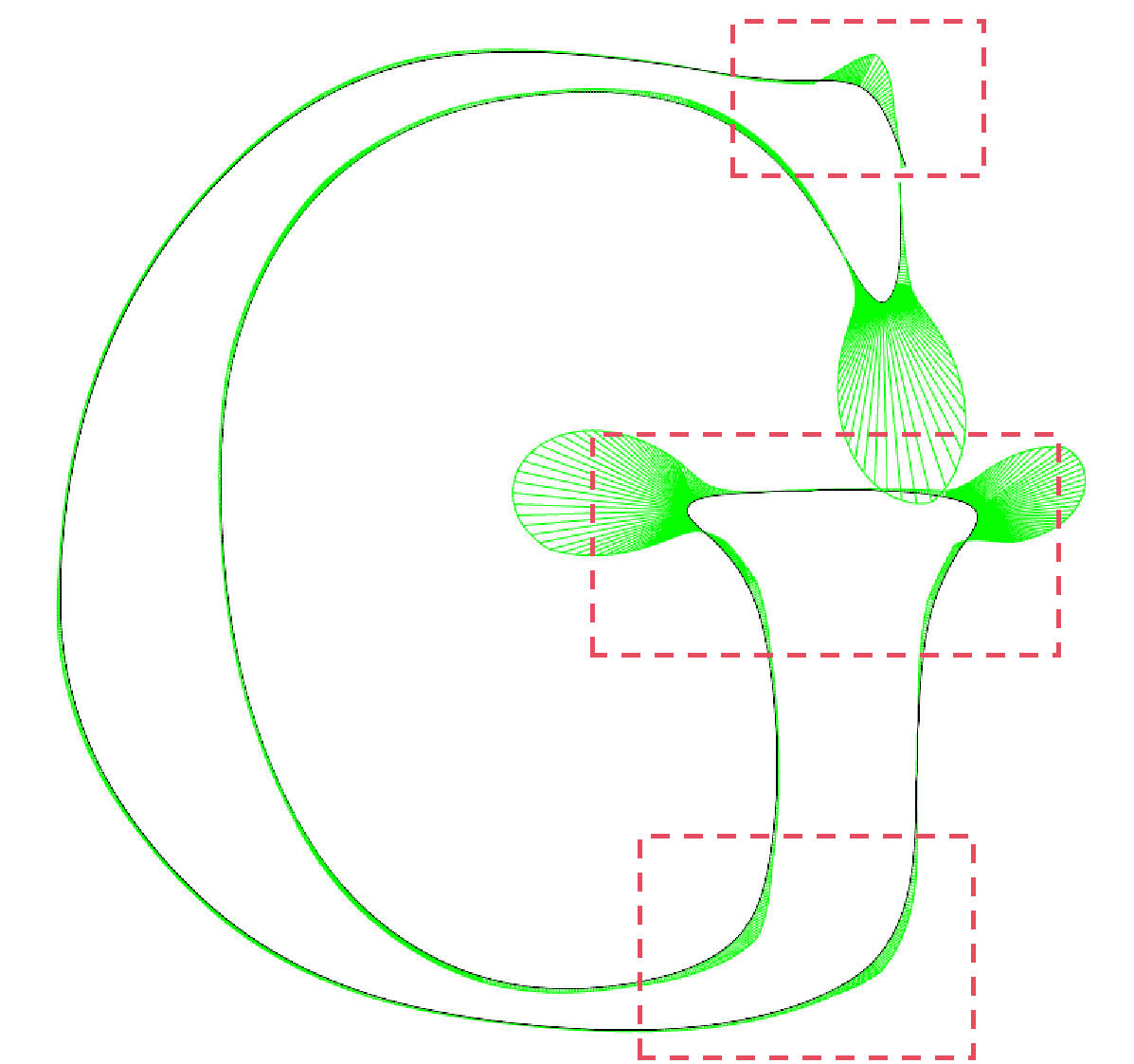}
	\end{minipage}
}
\subfigure[]{
	\begin{minipage}[b]{.3\linewidth}
		\centering
		\includegraphics[scale=0.22]{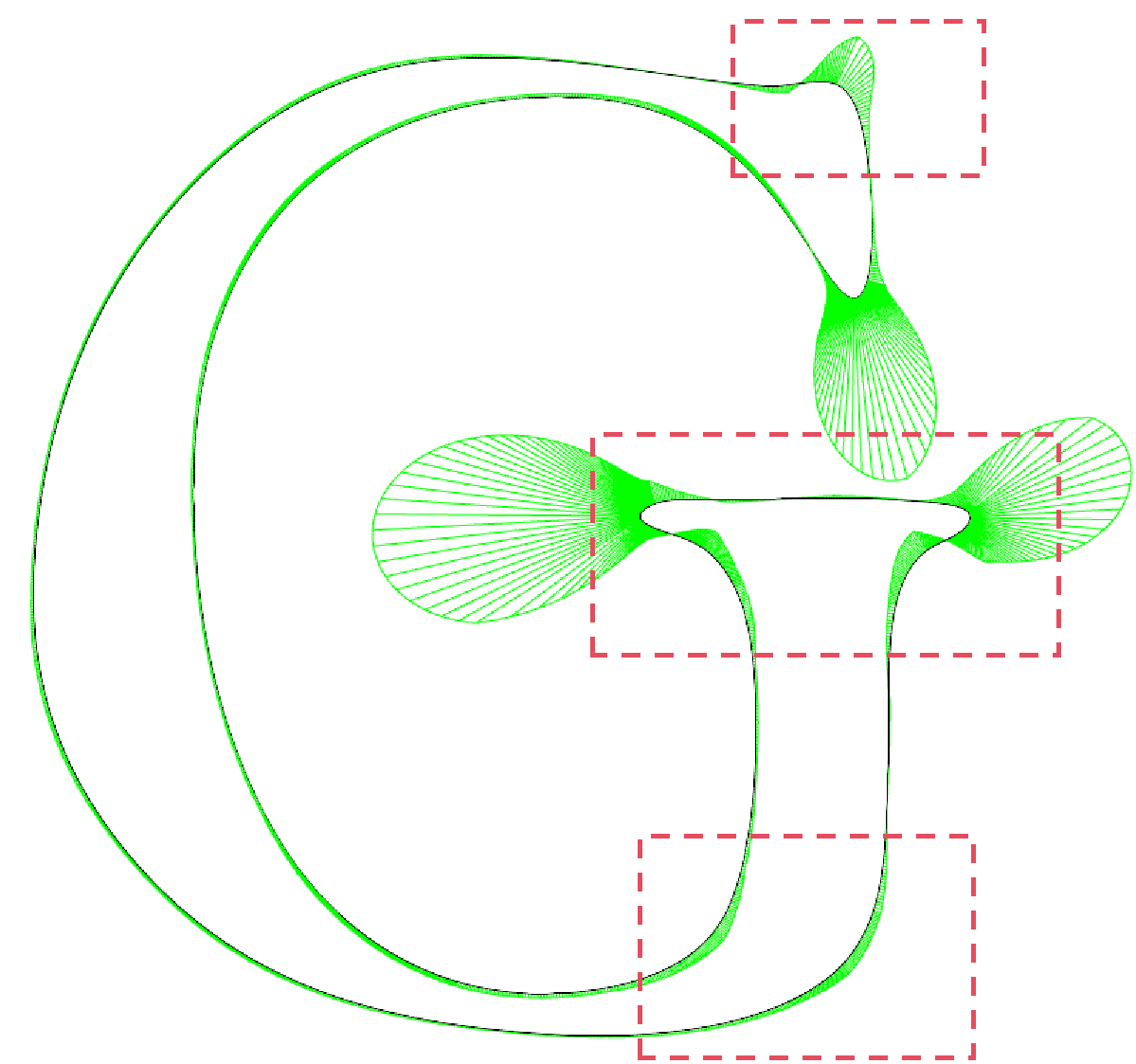}
	\end{minipage}
}

\caption{\label{fig2}Comparison results for the shape G model. We add the  curvature comb to show the fairness of the curves. (a) Original curve. (b) The curve fairing by the energy method. (c) The curve fairing by our method. }
\end{figure}

In the example of shape G, We demonstrated our algorithm's ability to make more detailed adjustments to the shape, which fairs the curve while still preserving its local sharp features. Figs.\ref{fig2}(a) is the  curve to be processed with the corresponding curvature comb. Set $\omega = 1\times 10^{-6}, r=2$,  and the curve can be faired by the energy methods(Figs.\ref{fig2}(b)). Though the curve is more fairing as a whole, the features in the first and second red box are also smoothed out, which is contrary to our original intention. Also, the fairness of the curve in the third red box needs to be further improved. In our algorithm, we set the fairing weights of the 1th to 5th and the 44th to 58th control points as  $1\times 10^{-7}$, and the 35th to 39th and the 63th to the 67th as $2\times 10^{-6}$ . The rest ones are remain as $1\times 10^{-6}$, so as to obtain a curve with good fairing effect and still retain local features(Figs.\ref{fig2}(c)). Our method which can preserve the   curve features is very practical and important.

\begin{figure}[H]
\centering
\subfigure[r=1, Iteration 0]{
	\begin{minipage}[b]{.2\linewidth}
		\centering
		\includegraphics[scale=0.22]{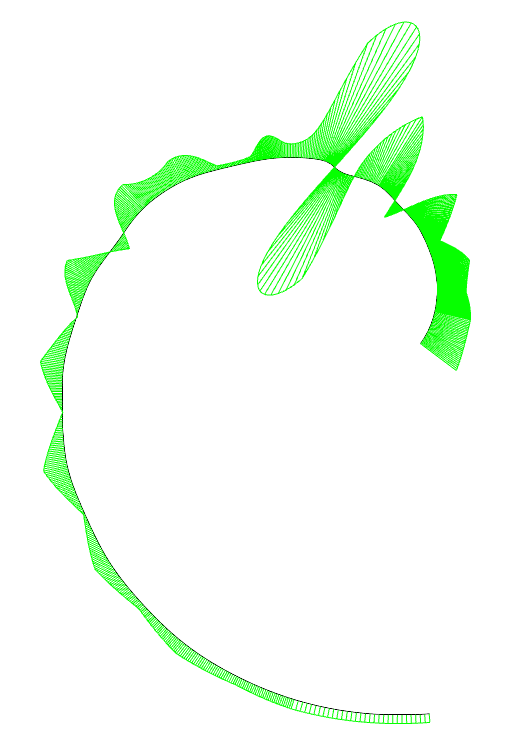}
	\end{minipage}
}
\subfigure[r=1, Iteration 5]{
	\begin{minipage}[b]{.2\linewidth}
		\centering
		\includegraphics[scale=0.22]{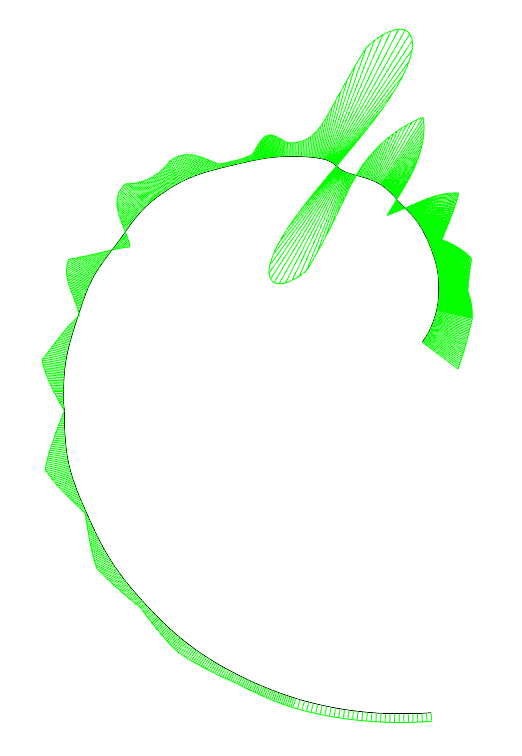}
	\end{minipage}
}
\subfigure[r=1, Iteration 10]{
	\begin{minipage}[b]{.2\linewidth}
		\centering
		\includegraphics[scale=0.22]{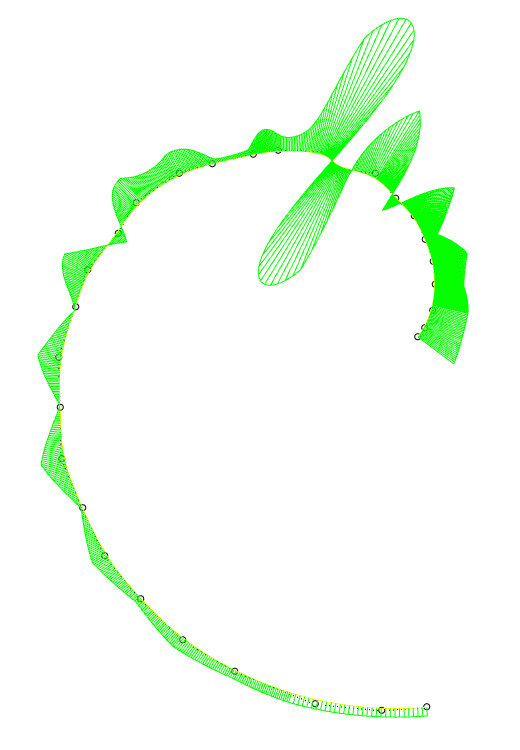}
	\end{minipage}
}
\subfigure[r=1, Iteration Stops]{
	\begin{minipage}[b]{.2\linewidth}
		\centering
		\includegraphics[scale=0.22]{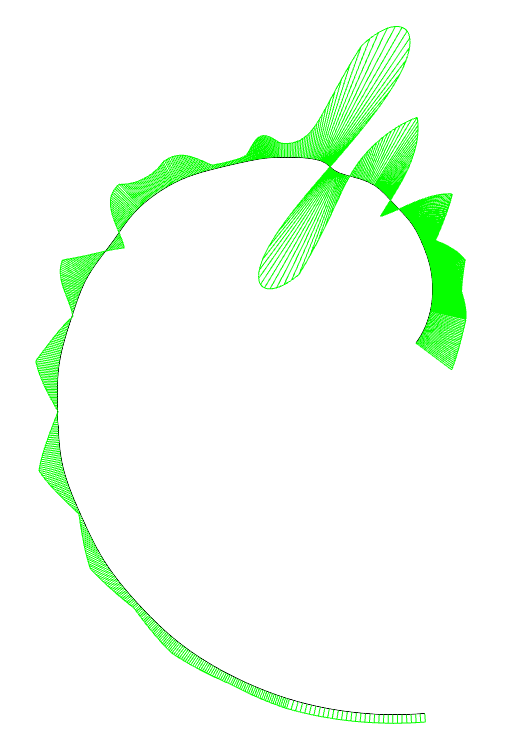}
	\end{minipage}
}

\subfigure[r=2, Iteration 0]{
	\begin{minipage}[b]{.2\linewidth}
		\centering
		\includegraphics[scale=0.22]{curve-fit.png}
		
	\end{minipage}
}
\subfigure[r=2, Iteration 5]{
	\begin{minipage}[b]{.2\linewidth}
		\centering
		\includegraphics[scale=0.22]{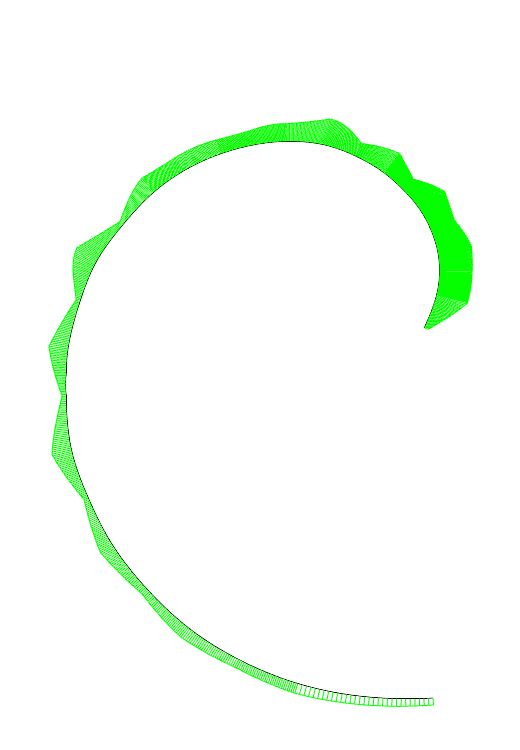}
		
	\end{minipage}
}
\subfigure[r=2, Iteration 10]{
	\begin{minipage}[b]{.2\linewidth}
		\centering
		\includegraphics[scale=0.22]{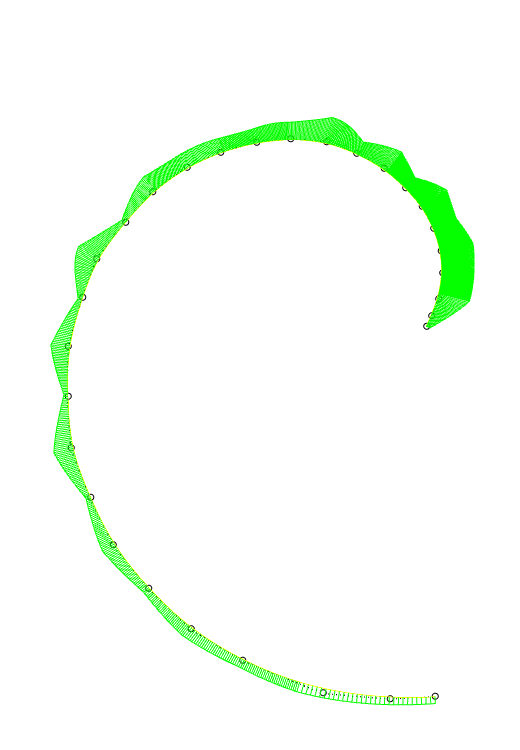}
		
	\end{minipage}
}
\subfigure[r=2, Iteration Stops]{
	\begin{minipage}[b]{.24\linewidth}
		\centering
		\includegraphics[scale=0.22]{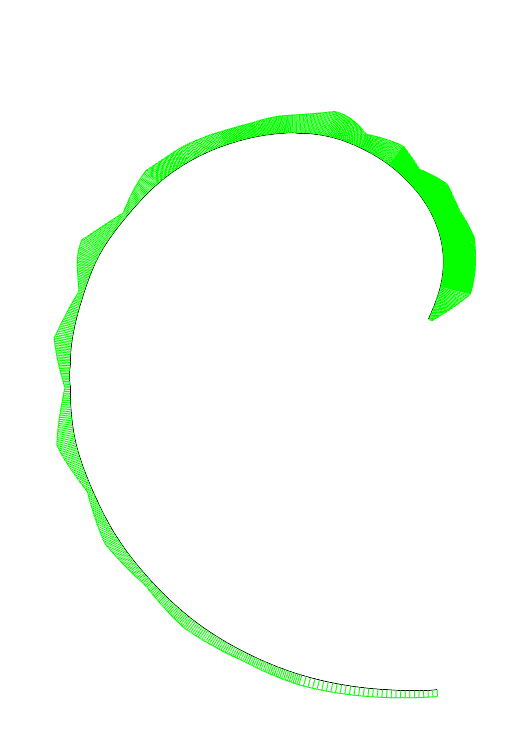}
		
	\end{minipage}
}
\subfigure[r=3, Iteration 0]{
	\begin{minipage}[b]{.2\linewidth}
		\centering
		\includegraphics[scale=0.22]{curve-fit.png}
		
	\end{minipage}
}
\subfigure[r=3, Iteration 5]{
	\begin{minipage}[b]{.2\linewidth}
		\centering
		\includegraphics[scale=0.22]{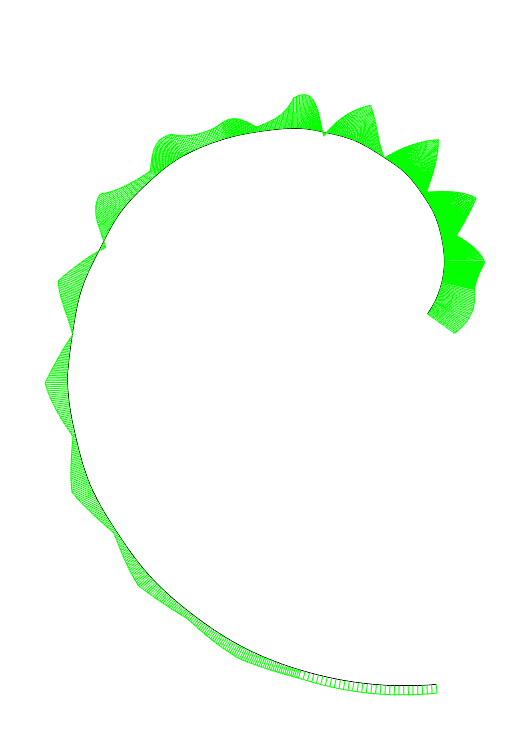}
		
	\end{minipage}
}
\subfigure[r=3, Iteration 10]{
	\begin{minipage}[b]{.2\linewidth}
		\centering
		\includegraphics[scale=0.22]{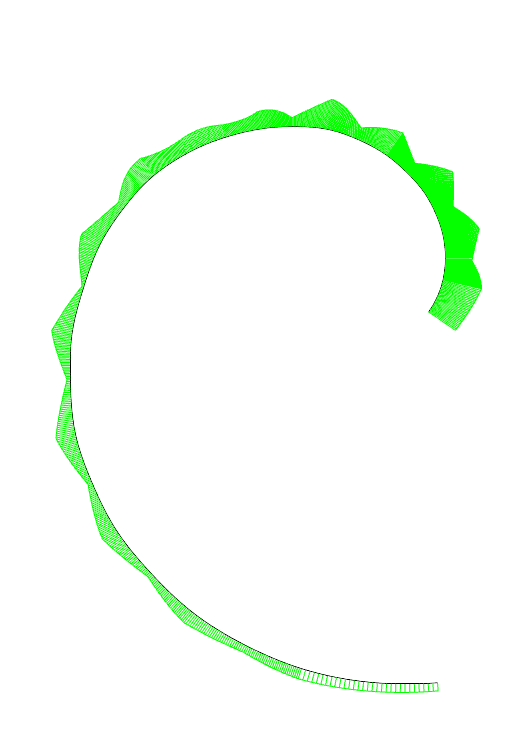}
		
	\end{minipage}
}
\subfigure[r=3, Iteration Stops]{
	\begin{minipage}[b]{.2\linewidth}
		\centering
		\includegraphics[scale=0.22]{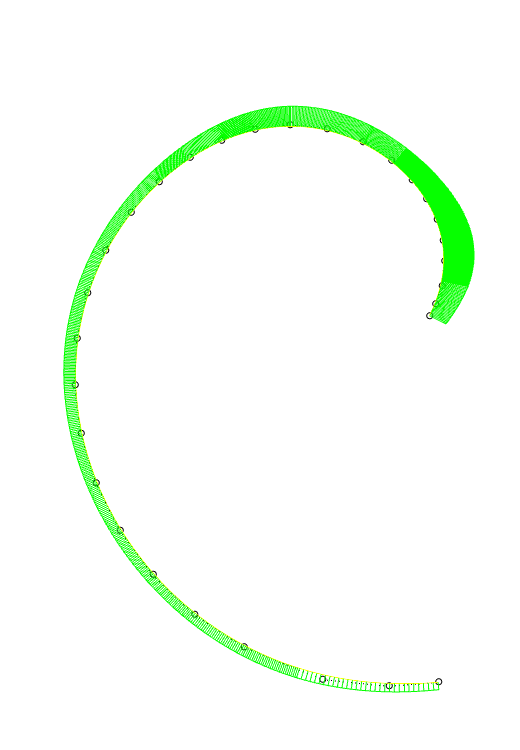}		
	\end{minipage}
}
\caption{\label{fig3}Comparison of different selections of $r$ (refer to Eq.(18)) in our method. Left-to-right: original curve, fairing curve after 5 iterations,  fairing curve after 10 iterations, fairing result. Top-to-bottom: fairing curves when $r=1$, $r=2$ and $r=3$. Curves are black, and the curvature comb is green.}
\end{figure}

In section 3.3, we mentioned the energy types corresponding to different values of $r$ (see Eq.(18)). Through numerical experiments, we compared the influence of different values of $r$ on the fairing effect of the curve (Figure \ref{fig3}) by using different fairing functionals. We randomly add Gaussian noise with 0 mean and 0.02 variances to the Archimedes helix as an initial curve: 

$$\left\{
\begin{aligned}
	x(t) & =  (2+1.5\theta)\cos\theta, \ \ \theta\in [0,5] \\
	y(t) & =  (2+1.5\theta)\sin\theta\\
\end{aligned}
\right.
$$

\begin{table}[h]
	\caption{Comparison of different selections of r.}\label{tab1}%
	\begin{tabular}{@{}ccccc@{}}
		\toprule
		Model  & RMSE  & Iteration  & Time  & Relative Energy(\%)\\ 
		\midrule
		Spiral(r=1)       &   0.009971          &       20      &     0.0208773         &      98.25\% \\
		Spiral(r=2)      &   0.069380              &    120         &    0.0209216       &    42.41\%     \\
		Spiral(r=3)      &     0.230256           &       $>$800      &      0.0219848     &     4.25\% \\
		\botrule
	\end{tabular}
	
\end{table}


\begin{table}[h]
\tabcolsep=0.4cm
\caption{\label{tab2}Statistical data of our method on curve and surface examples.}
\begin{tabular}{ccccc}
		\hline
		\multirow{2}{*}{Model} & \multirow{2}{*}{\#Control Points} & \multirow{2}{*}{\#Iteration} & \multicolumn{2}{c}{Running Time (s)} \\ 
		&                       &                              & Energy Method      & Our Method      \\ \hline
		Dolphin              & 60                      &   130                           &    0.140523             &   0.131524              \\
		Shape G              & 100                      &  114                            
		&   0.561234         &     0.542875     \\
		Spiral(r=1)          & 30                       &    20                          &    0.020921            &         0.020065        \\
		Spiral(r=2)          & 30                      &   120                     
		&    0.020877       &    0.021396           \\
		Spiral(r=3)          & 30                      &    $>800$                          &     0.021568               &   0.023015            \\
		Airfoil Surface      &       $42\times 21$               &    161              &     15.42069               &   15.46335             \\
		Phone surface       &      $8\times 36$                &        421          &   
		0.26118           &     0.21747          \\
		Car surface           &       $16\times 22$               &     867             &      1.42560             &       1.38619    \\ \hline
	\end{tabular}
\end{table}

\begin{table}[h]
\tabcolsep=0.5cm
\caption{\label{tab3}Comparison of root mean squared distance errors and energy.}
	\begin{tabular}{ccccc}
		\hline
		& \multicolumn{2}{c}{Traditional Energy Method}  & \multicolumn{2}{c}{Our Method}    \\ \hline
		Model  & RMSE  & Absolute Energy  & RMSE  & Absolute Energy\\
		Dolphin           &  0.89524 & $4.05743\times10^7$ & 1.55481 & $1.47889\times10^7$  \\
		Shape G           &  0.08945 &   14551.6          & 0.00715  &  26162.9      \\
		Airfoil Surface   & 2.43569 &  $3.69854\times 10^7$& 3.03233   & $2.21999\times10^7$ \\
		Phone Surface    &    0.01443  &    3.06314 &  0.01549     &  3.0488       \\
		Car Surface      &     1.54137    &    1.45931     &    1.96307   &  1.30684         \\ \hline
	\end{tabular}

\end{table}

\begin{figure}[H]
\centering
\subfigure[]{
	\begin{minipage}[b]{.4\linewidth}
		\centering
		\includegraphics[scale=0.53]{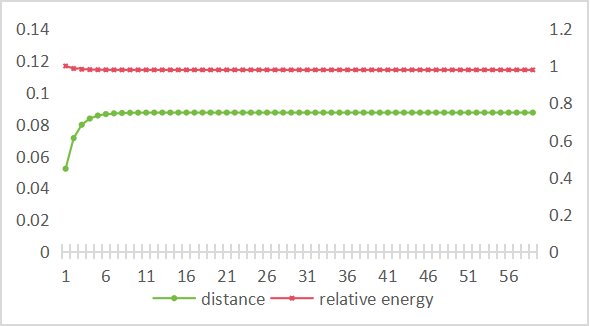}
	\end{minipage}
}
\subfigure[]{
	\begin{minipage}[b]{.4\linewidth}
		\centering
		\includegraphics[scale=0.53]{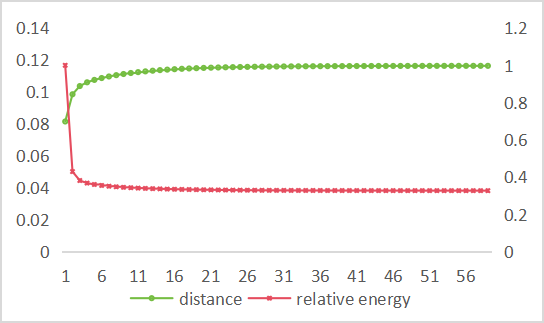}
	\end{minipage}
}
\subfigure[]{
	\begin{minipage}[b]{.4\linewidth}
		\centering
		\includegraphics[scale=0.53]{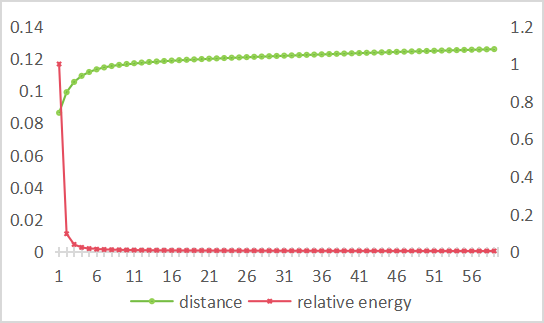}
	\end{minipage}
}

\caption{\label{fig4}Diagrams of iteration v.s. relative iterative error and relative energy of the Archimedean spiral in our method. (a) r=1. (b) r=2. (c) r=3.}
\end{figure}

Also, the algorithm iteration procedures are illustrated in Figure \ref{fig4} for $r=1,2,3$. As shown in Figure \ref{fig4}, the fairness of curves is improves with iterations, and the fairness for $r=3$ is the best while the fairness for $r=1$ is the worst. In addition, the diagrams of relative energy and control points' deviation are presented in Figure \ref{fig4}. The energy of curves rapidly decreases while the deviation of control points increases in the early iterations. In addition, it is obvious that the algorithm  converges fastest when $r=1$ and converges slowest when $r=3$. Detailed data are shown in Table \ref{tab1}.

The statistic of our method on the curve examples are listed in Table \ref{tab2} and the deviation of control points and absolute energies are shown in Table \ref{tab3}. According to Table \ref{tab3}, the algorithm proposed in this paper can effectively fair the curve while minimizing the deviation of control points. In addition, compared with the traditional energy-based method, the proposed algorithm is more flexible and can adjust the curve shape with higher precision.

\subsection{Surface}
In this section, the effectiveness of the algorithm is demonstrated by applying it to five surface models: an Airfoil, a Phone, a Car, a Mannequin, and a Twisted Cuboid. The fairness of the resulting surfaces is evaluated by visualizing absolute mean curvature distributions and zebra maps. The statistic of the method on the first three surface examples are listed in Table \ref{tab2}.

\begin{figure}[H]
\centering
\subfigure[]{
	\begin{minipage}[b]{.3\linewidth}
		\centering
		\includegraphics[scale=0.23]{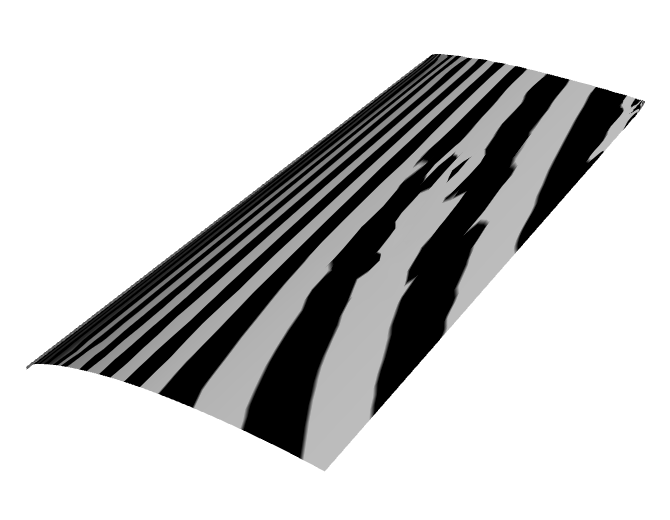}
	\end{minipage}
}
\subfigure[]{
	\begin{minipage}[b]{.3\linewidth}
		\centering
		\includegraphics[scale=0.23]{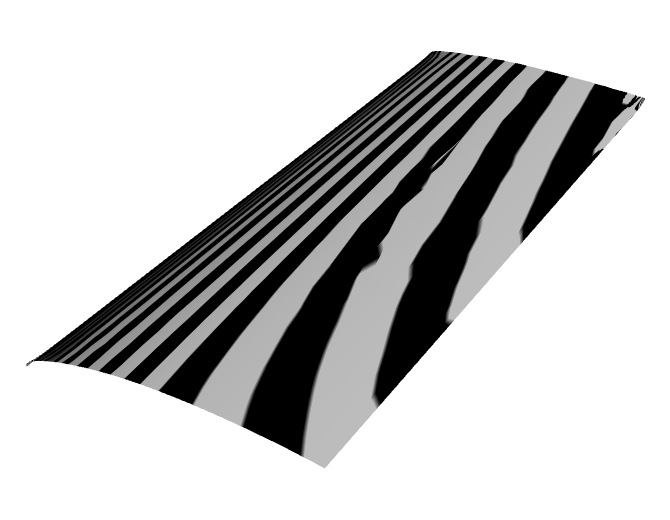}
	\end{minipage}
}
\subfigure[]{
	\begin{minipage}[b]{.3\linewidth}
		\centering
		\includegraphics[scale=0.23]{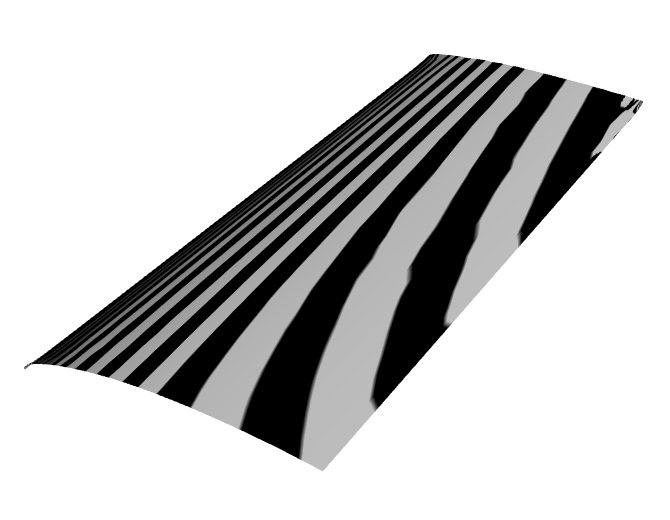}
	\end{minipage}
}
\subfigure[]{
	\begin{minipage}[b]{.3\linewidth}
		\centering
		\includegraphics[scale=0.23]{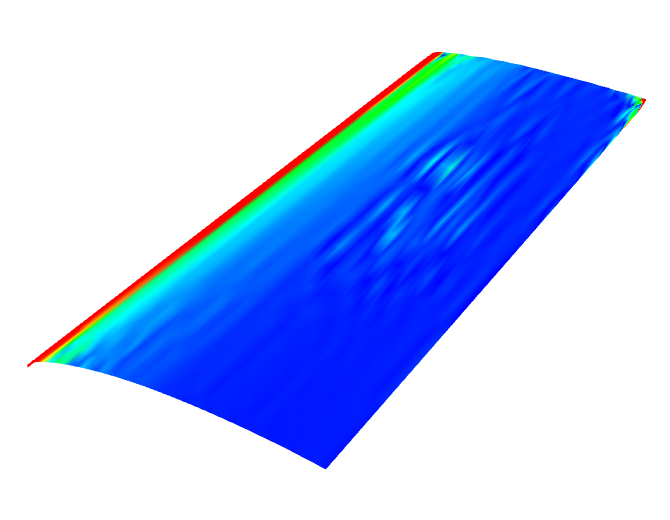}
	\end{minipage}
}
\subfigure[]{
	\begin{minipage}[b]{.3\linewidth}
		\centering
		\includegraphics[scale=0.23]{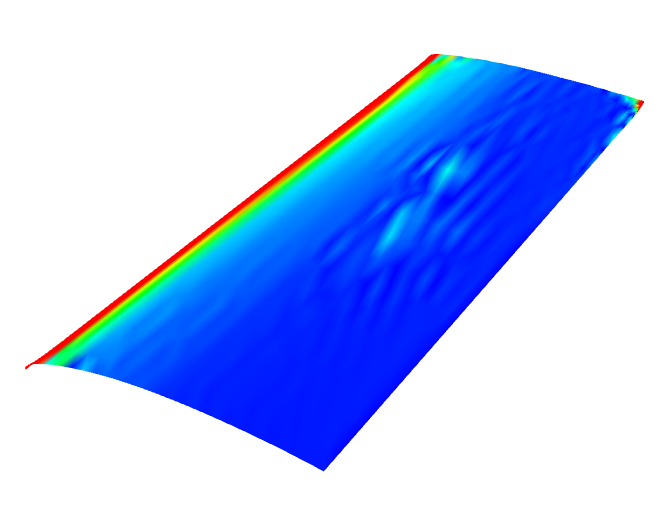}
	\end{minipage}
}
\subfigure[]{
	\begin{minipage}[b]{.3\linewidth}
		\centering
		\includegraphics[scale=0.23]{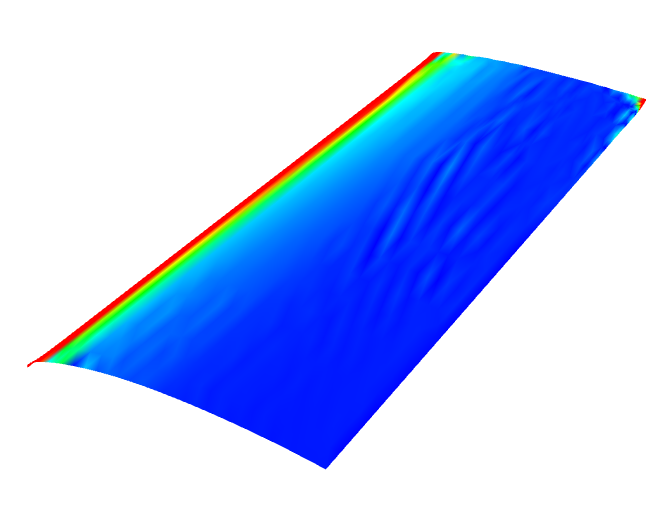}
	\end{minipage}
}
\caption{\label{fig5}Comparison results for the airfoil surface.  (a)(d) Original Surface. (b)(e) The surface fairing by the energy method. (c)(f) The surface fairing by our method. (a)(b)(c)Zebra diagram of surfaces. (d)(e)(f) Absolute value of mean curvature on the surfaces.}
\end{figure}

\begin{figure}[H]
	\centering
	\subfigure[]{
		\begin{minipage}[b]{.47\linewidth}
			\centering
			\includegraphics[scale=0.2]{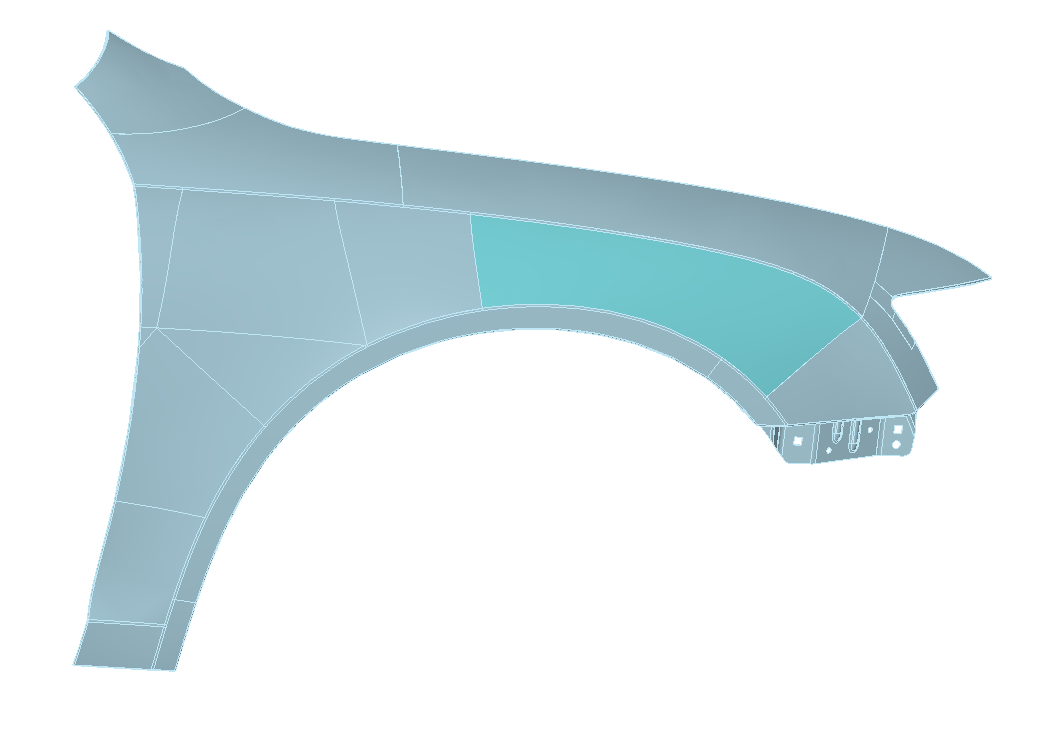}
		\end{minipage}
	}
	\subfigure[]{
		\begin{minipage}[b]{.47\linewidth}
			\centering
			\includegraphics[scale=0.2]{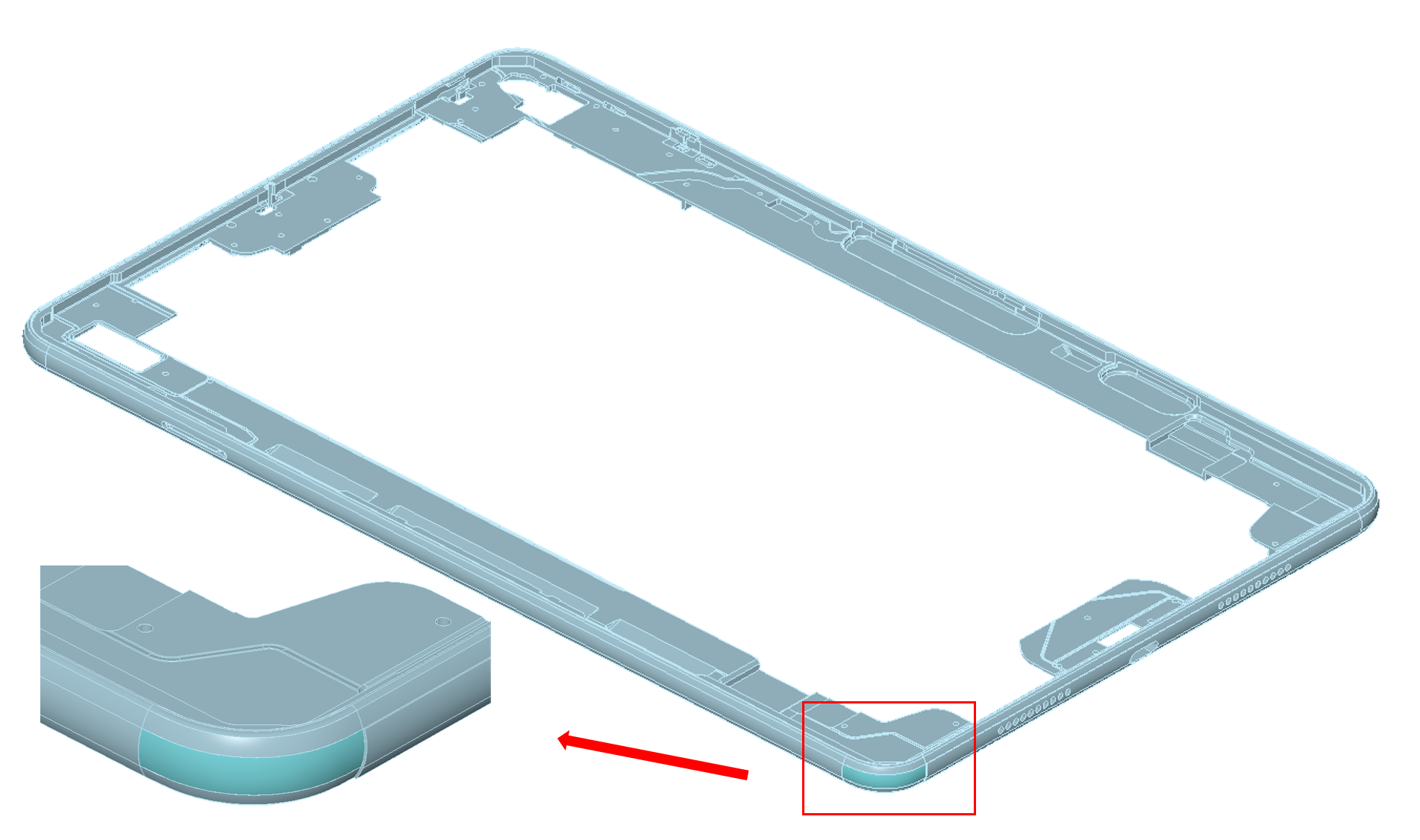}
		\end{minipage}
	}
	\caption{\label{fig6}Two surface models. (a) Car model. The highlighted surface is the model in Figure \ref{fig7}; (b) Phone model. The highlighted surface is the model in Figure \ref{fig8}.}
\end{figure}

For the airfoil surface, we take the fairing weight $\omega$ as $2\times 10^{-4}$ in the traditional energy minization method. The overall fairness of the surface has been further improved, but the local fairness is still not good enough (Figs.\ref{fig5}(b), Figs.\ref{fig5}(e)). Some fairing weights are increased to $0.002$ in our method to improve the fairness of some regions, whereas the others remain $2\times 10^{-4}$. Our method enables more fine-grained and flexible manipulation of the surface shape, resulting in a fairer result (Figs.\ref{fig5}(c), Figs.\ref{fig5}(f)).

\begin{figure}[H]
	\centering
	\subfigure[]{
		\begin{minipage}[b]{.3\linewidth}
			\centering
			\includegraphics[scale=0.14]{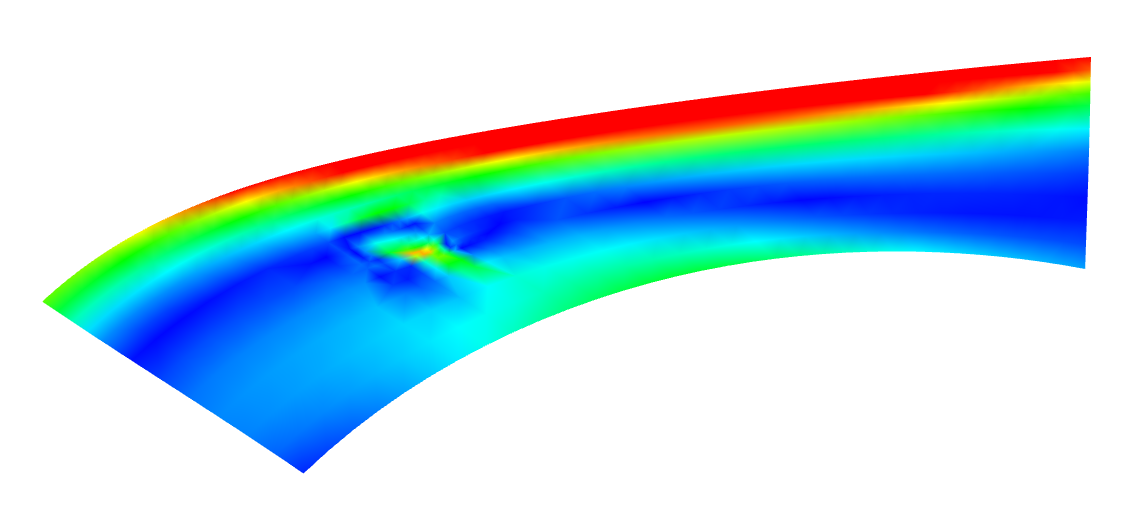}
		\end{minipage}
	}
	\subfigure[]{
		\begin{minipage}[b]{.3\linewidth}
			\centering
			\includegraphics[scale=0.14]{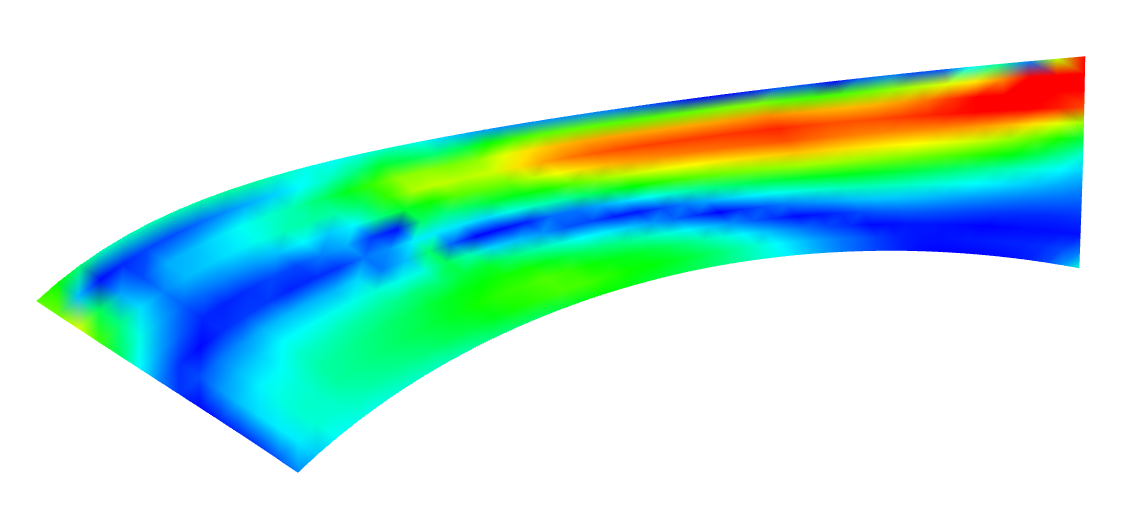}
		\end{minipage}
	}
	\subfigure[]{
		\begin{minipage}[b]{.3\linewidth}
			\centering
			\includegraphics[scale=0.14]{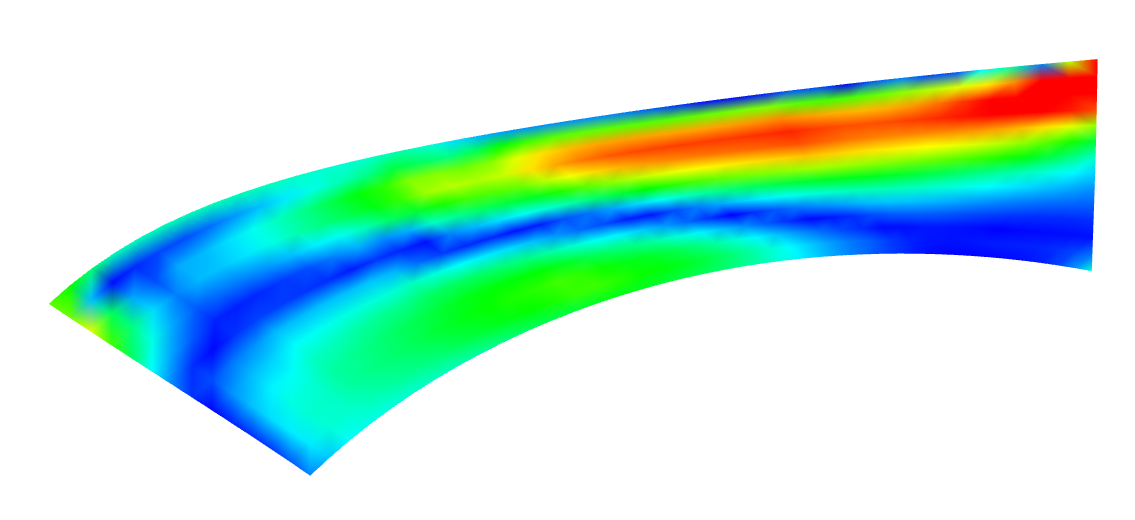}
		\end{minipage}
	}
	\subfigure[]{
		\begin{minipage}[b]{.3\linewidth}
			\centering
			\includegraphics[scale=0.14]{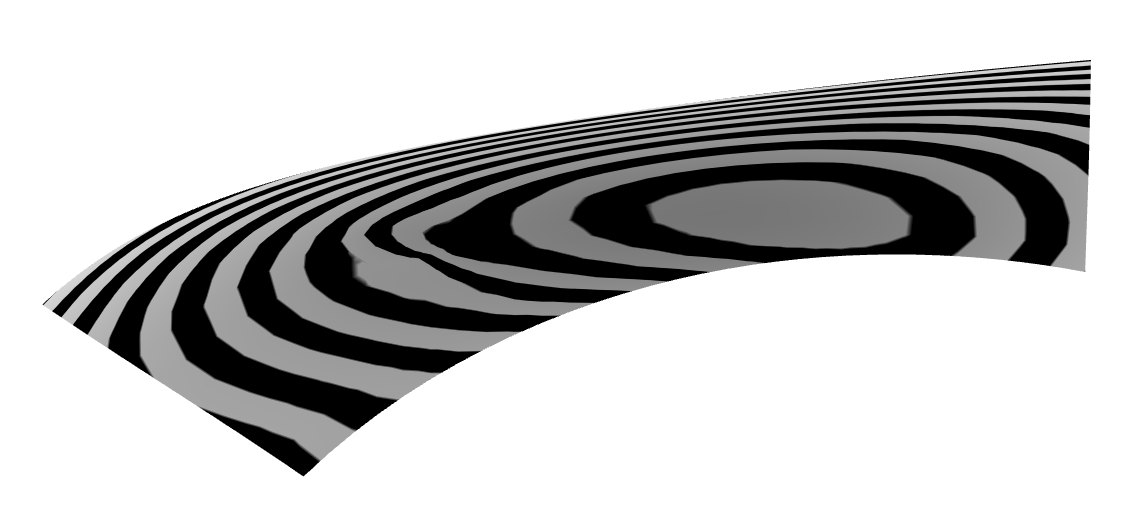}
		\end{minipage}
	}
	\subfigure[]{
		\begin{minipage}[b]{.3\linewidth}
			\centering
			\includegraphics[scale=0.14]{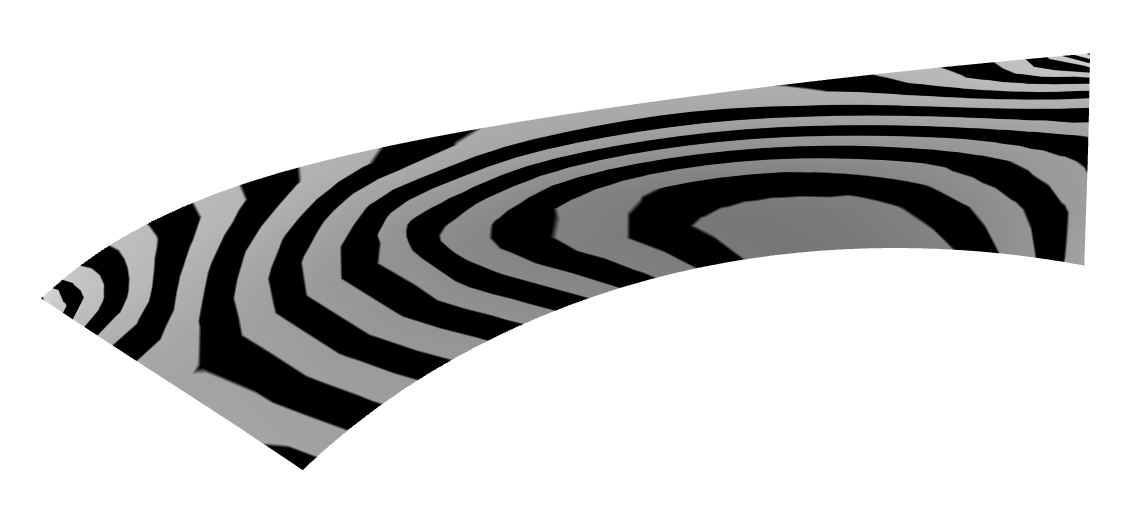}
		\end{minipage}
	}
	\subfigure[]{
		\begin{minipage}[b]{.3\linewidth}
			\centering
			\includegraphics[scale=0.14]{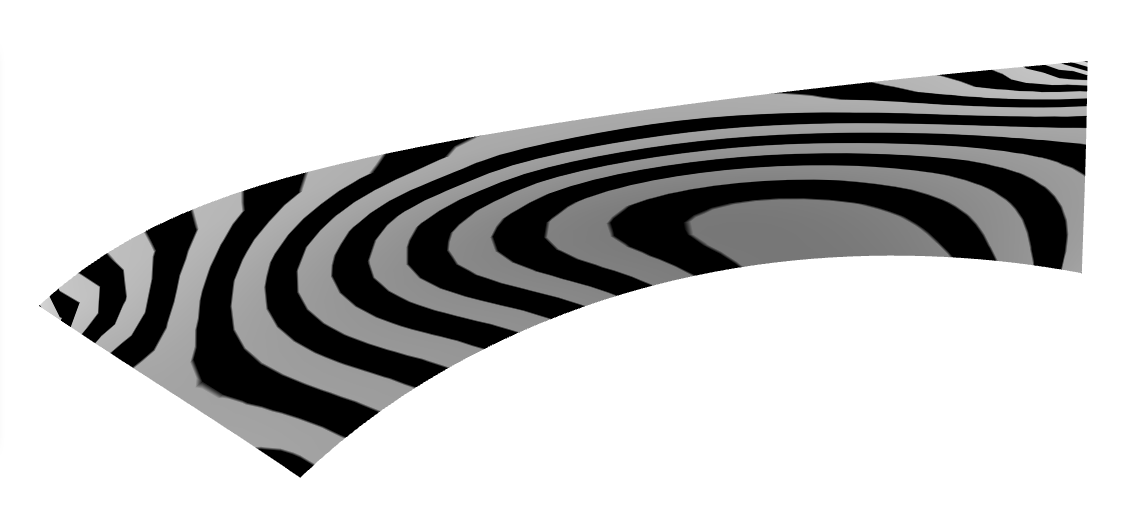}
		\end{minipage}
	}
	\caption{\label{fig7}Comparison results for the Car model. (a)(d) Original Surface. (b)(e) The surface fairing by the energy method. (c)(f) The surface fairing by our method. (a)(b)(c) Absolute value of mean curvature on the surfaces. (d)(e)(f) Zebra diagram on surfaces.}
\end{figure}

For the car model, the fairing weight $\omega$ is taken as $0.1$ in the traditional energy minimization mothod. The overall fairness of the surface has been improved, but there are still some areas with poor fairness locally. Traditional fairing methods can only enhance the fairness by adjusting the surface globally. In our method, to fair the surface in this model, we set the fairing weights corresponding to the region with relatively poor local fairness to 0.18, and set the weights of the remaining areas at 0.1. In this case, we can obtain a more fairing model(Figs.\ref{fig7}(c), Figs.\ref{fig7}(f)).

\begin{figure}[H]
\centering
\subfigure[]{
	\begin{minipage}[b]{.3\linewidth}
		\centering
		\includegraphics[scale=0.28]{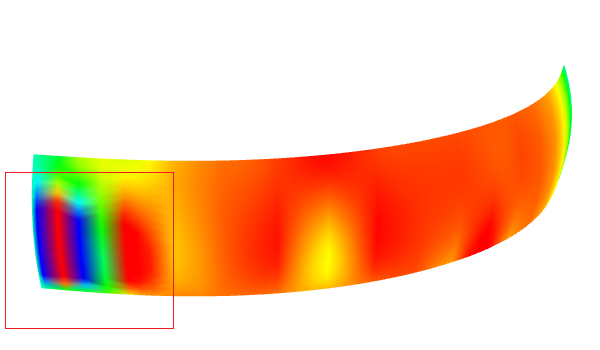}
	\end{minipage}
}
\subfigure[]{
	\begin{minipage}[b]{.3\linewidth}
		\centering
		\includegraphics[scale=0.28]{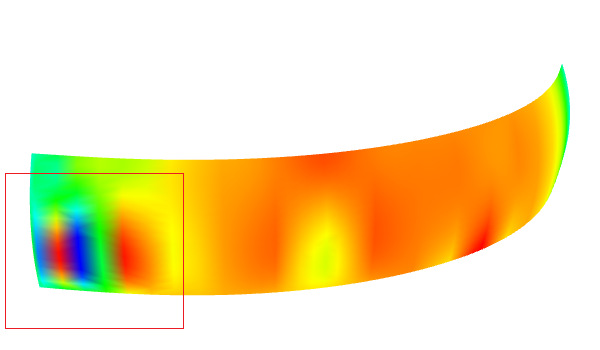}
	\end{minipage}
}
\subfigure[]{
	\begin{minipage}[b]{.3\linewidth}
		\centering
		\includegraphics[scale=0.28]{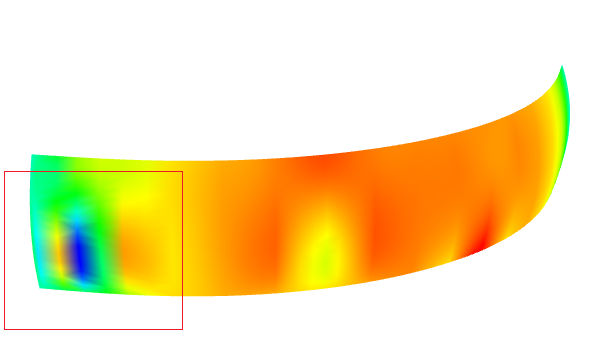}
	\end{minipage}
}
\subfigure[]{
	\begin{minipage}[b]{.3\linewidth}
		\centering
		\includegraphics[scale=0.28]{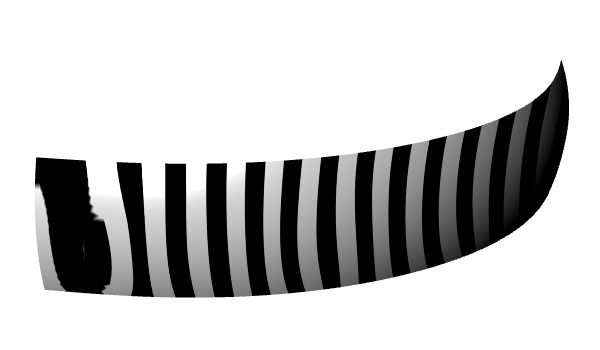}
	\end{minipage}
}
\subfigure[]{
	\begin{minipage}[b]{.3\linewidth}
		\centering
		\includegraphics[scale=0.28]{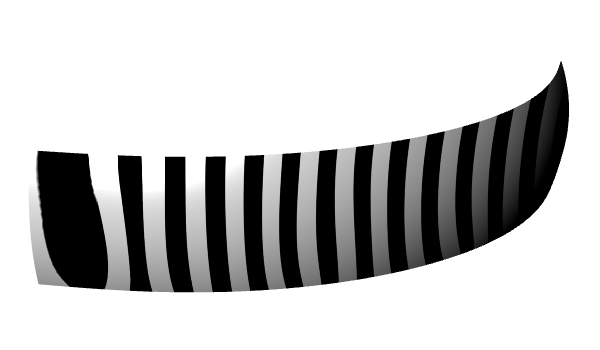}
	\end{minipage}
}
\subfigure[]{
	\begin{minipage}[b]{.3\linewidth}
		\centering
		\includegraphics[scale=0.28]{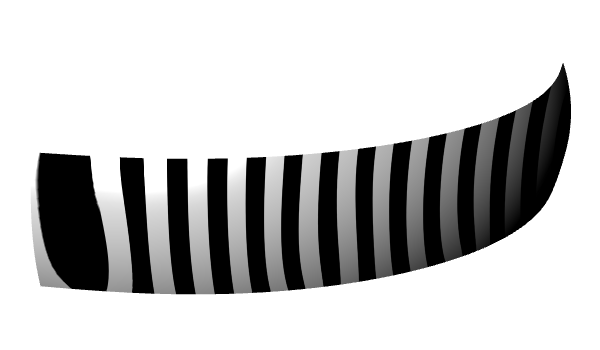}
	\end{minipage}
}
\caption{\label{fig8}Comparison results for the Phone model. (a)(d) Original Surface. (b)(e) The surface fairing by the energy method. (c)(f) The surface fairing by our method. (a)(b)(c) Absolute value of mean curvature on the surfaces. (d)(e)(f) Zebra diagram on surfaces. }
\end{figure}

\begin{figure}[H]
	\centering
	\subfigure[]{
		\begin{minipage}[b]{.42\linewidth}
			\centering
			\includegraphics[scale=0.3]{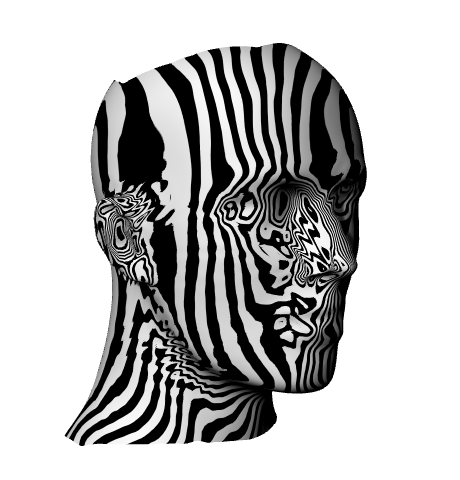}
		\end{minipage}
	}
	\subfigure[]{
		\begin{minipage}[b]{.42\linewidth}
			\centering
			\includegraphics[scale=0.3]{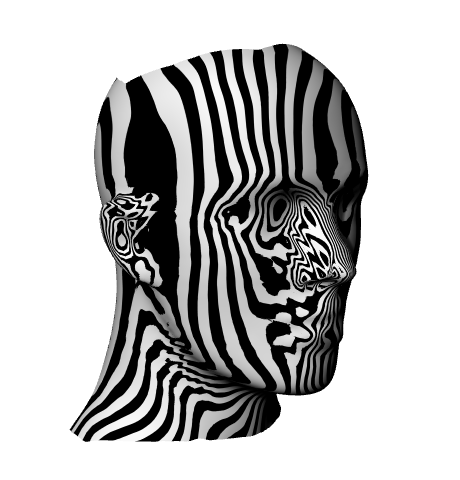}
		\end{minipage}
	}
	\subfigure[]{
		\begin{minipage}[b]{.42\linewidth}
			\centering
			\includegraphics[scale=0.2]{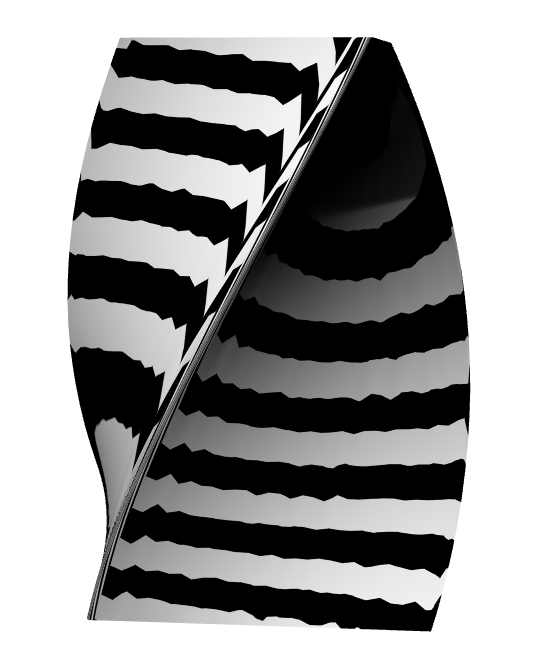}
		\end{minipage}
	}
	\subfigure[]{
		\begin{minipage}[b]{.42\linewidth}
			\centering
			\includegraphics[scale=0.2]{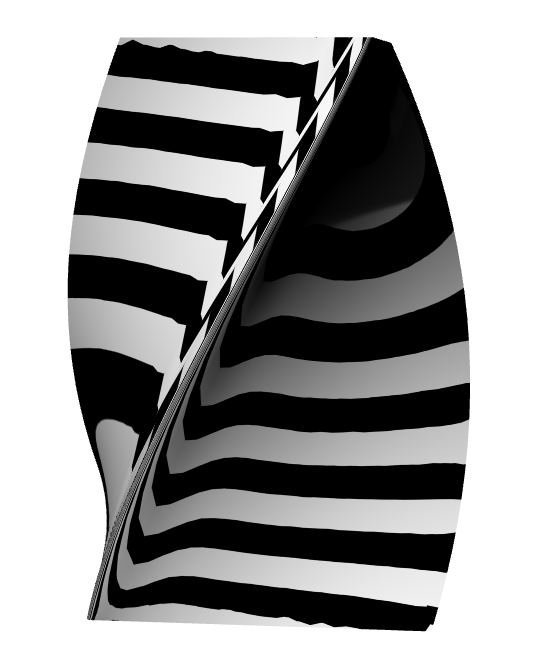}
		\end{minipage}
	}
	\caption{\label{fig9}Fairing results for the Mannnequin model and the twisted cuboid model. (a) Original Mannnequin model. (b) The Mannnequin model fairing by our method. (c) Original twisted cuboid  model. (d) The twisted cuboid model fairing by our method. }
\end{figure}

The fairing weight for the car model in the traditional energy minization method is taken as $\omega = 5\times 10^{-4}$(Figs.\ref{fig8}(b), Figs.\ref{fig8}(e)). Similarily, the corresponding fairing weights are increased to $1\times 10^{-3}$ in our method, and the other fairing weights are set as $5\times 10^{-4}$ to further improve the fairness of the surface(Figs.\ref{fig8}(c), Figs.\ref{fig8}(f)).

Also, we use the Mannequin model and the twisted cuboid model to show the ability of our algorithm for fine adjustment of surfaces. In the Mannequin model, we set the fairing weights corresponding to the nose as $1\times 10^{-7}$, and the other fairing weights as  $5\times 10^{-4}$. And in the twisted cuboid model, the fairing weights corresponding to the edges are set as  $1\times 10^{-7}$ and the others are set as  $3\times 10^{-4}$. In this two cases, we can obtain fairing models while retaining local sharp features(Figure \ref{fig9}).

\subsection{Fairing Method with Automatic selection of Control Points}


This section presents examples of applying the automatic fairing algorithm described in Section 3.4 to adjust curves and surfaces. The algorithm facilitates the appropriate selection of control points for adjustment and can perform fairing automatically. Additionally, fairing weights can be assigned based on the sorted order of the control points.

\begin{figure}[H]
	\centering
	\subfigure[]{
		\begin{minipage}[b]{.4\linewidth}
			\centering
			\includegraphics[scale=0.2]{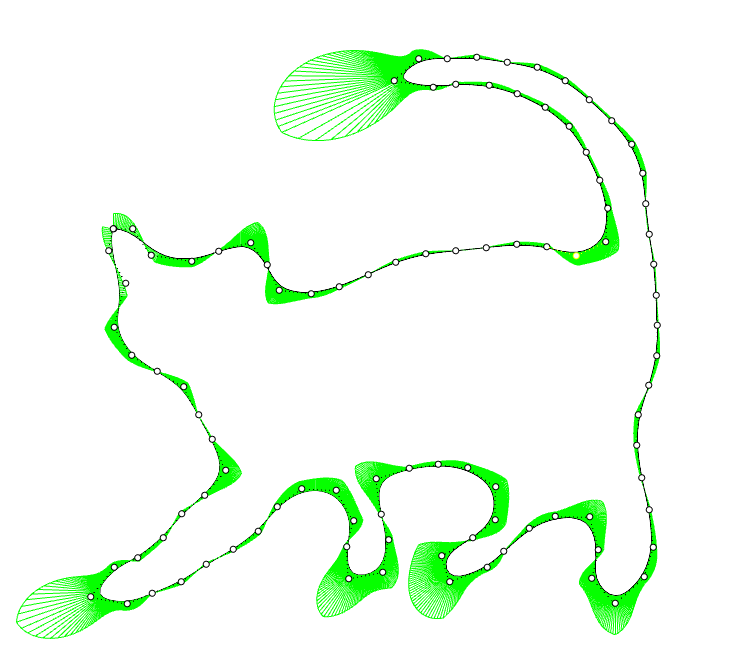}
		\end{minipage}
	}
	\subfigure[]{
		\begin{minipage}[b]{.4\linewidth}
			\centering
			\includegraphics[scale=0.2]{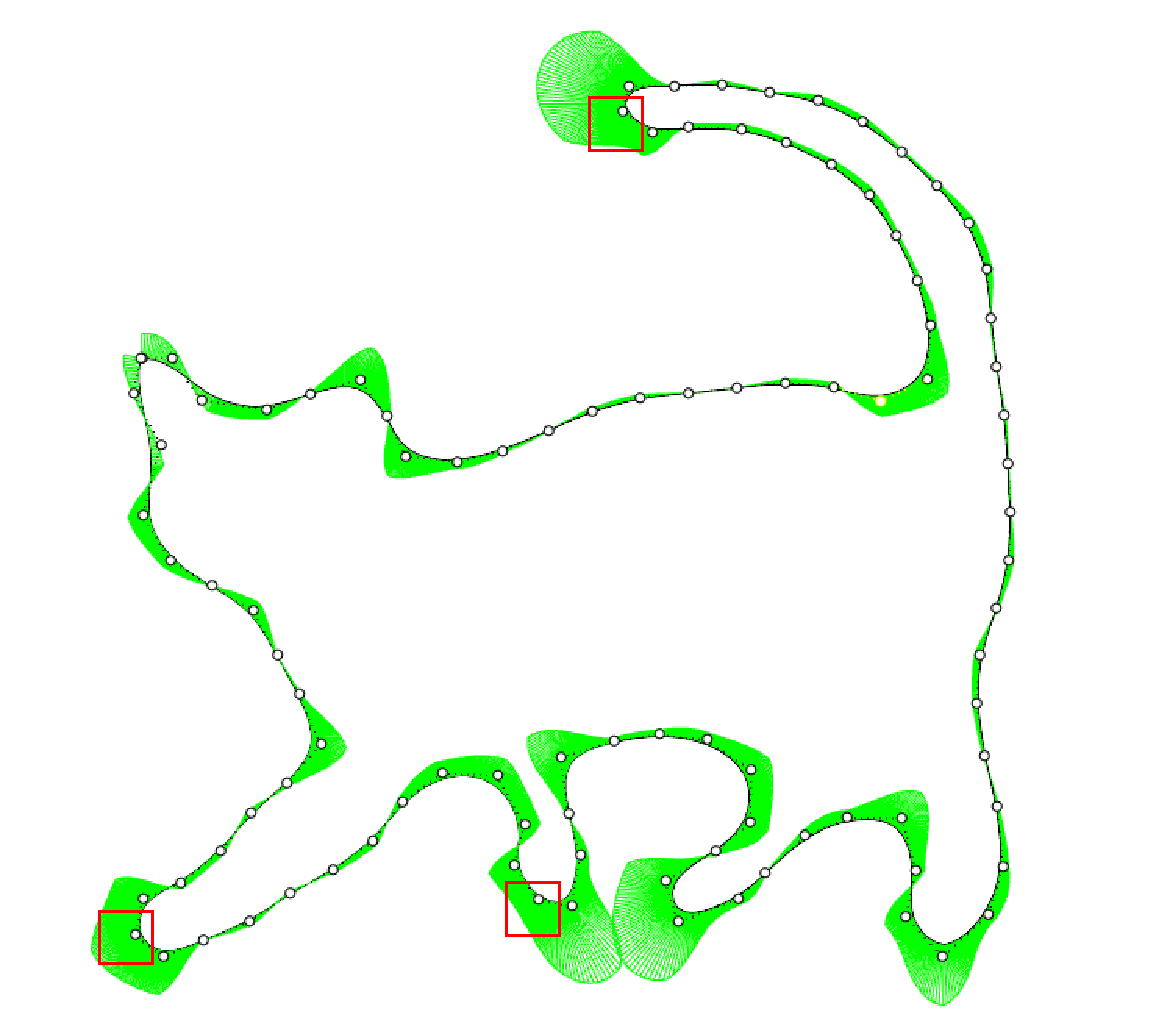}
		\end{minipage}
	}
	\subfigure[]{
		\begin{minipage}[b]{.4\linewidth}
			\centering
			\includegraphics[scale=0.2]{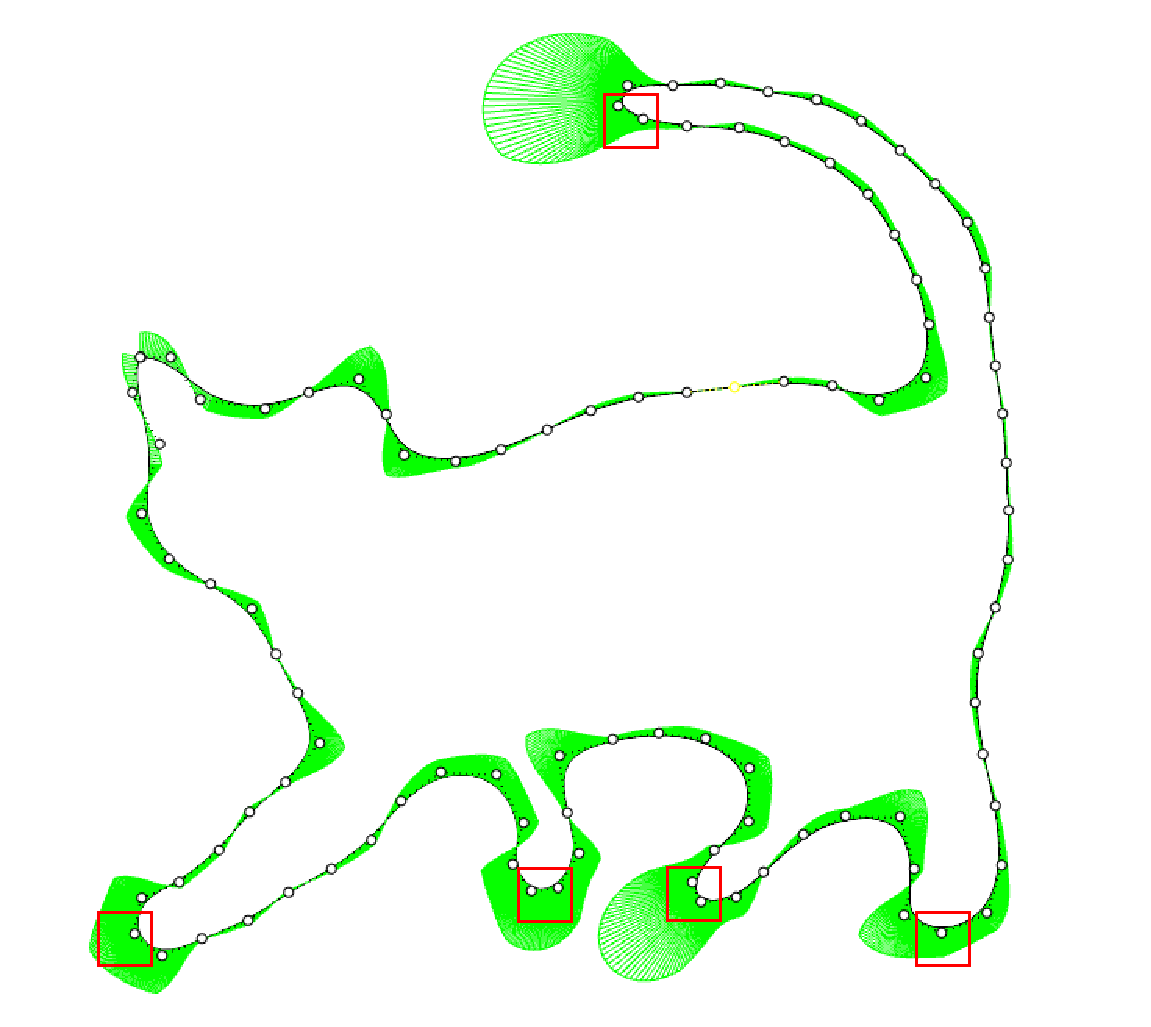}
		\end{minipage}
	}
	\subfigure[]{
		\begin{minipage}[b]{.4\linewidth}
			\centering
			\includegraphics[scale=0.2]{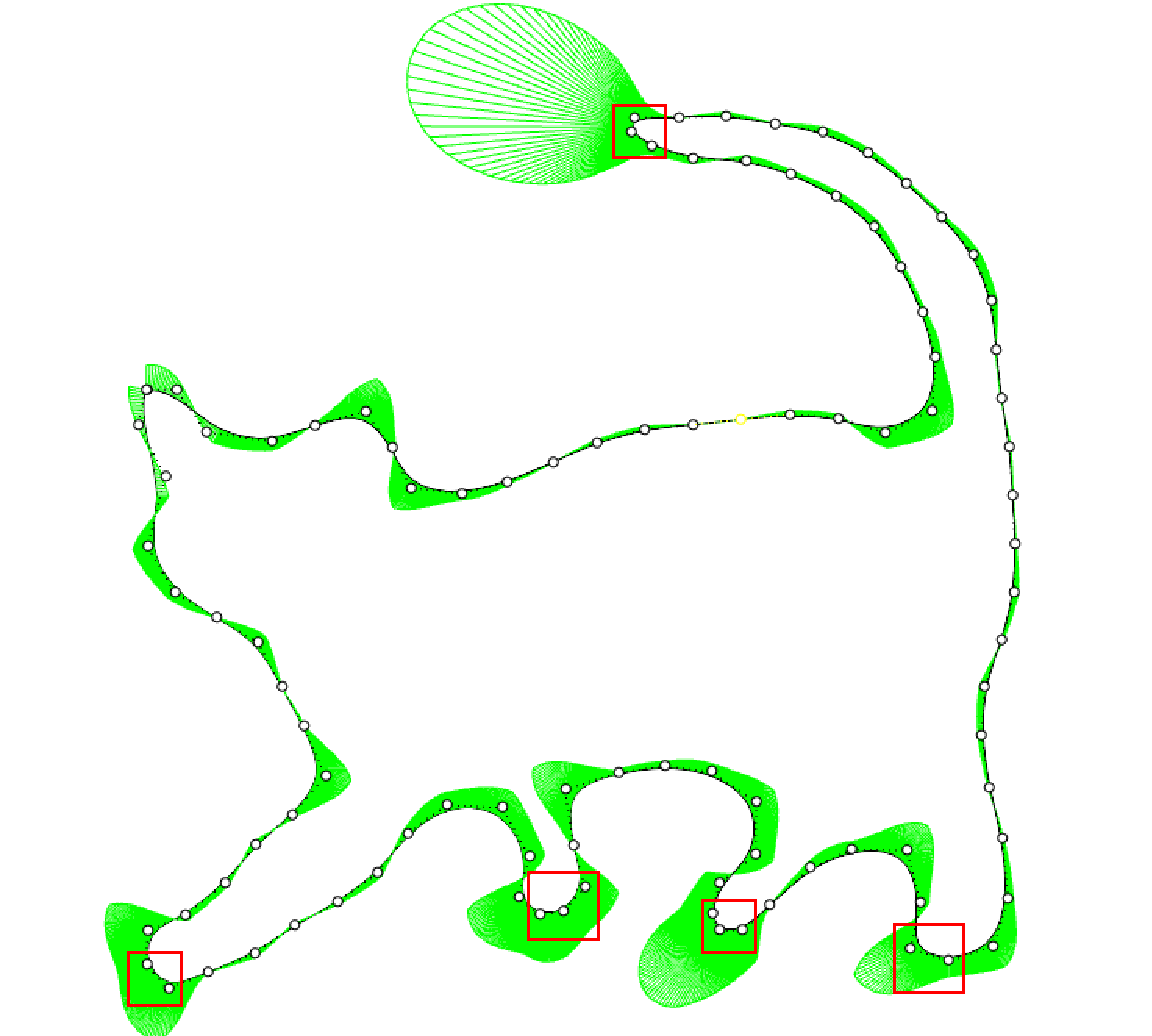}
		\end{minipage}
	}
	\caption{\label{fig10}Fairing results for the Cat model. (a) Original curve. (b)  The first local fairing by adjusting 3 selected control points in the red boxes. (c)The second local fairing by adjusting 8 selected control points in the red boxes. (d) The third local fairing by adjusting the fairing weights of the same 8 control points in Figs.\ref{fig10}.(c).}
\end{figure}

Figure \ref{fig10} is a cubic B-spline curve with 100 control points. Figs.\ref{fig10}(a) shows the original curve and its curvature comb. We select to adjust 3 control points(this can also be modified to use a certain percentage as the adjustment target in practical applications), and set the fairing weight to $1\times 10^{-6}$. Figs.\ref{fig10}(b) presents the faired curve, where the control points within the red boxes are the combination of control points to be adjusted, selected by the automatic algorithm. If you want the curve to undergo more adjustments, select 8 control points for adjustment while keeping the fairing weight unchanged;  Figs.\ref{fig10}(c) shows the result after fairing. We can also set the fairing weights more precisely: for example, set the first 4 fairing weights to $5\times 10^{-5}$ and the last 4 to $1\times 10^{-6}$, with the faired result shown in Figs.\ref{fig10}(d).

\begin{figure}[H]
	\centering
	\subfigure[]{
		\begin{minipage}[b]{.45\linewidth}
			\centering
			\includegraphics[scale=0.1]{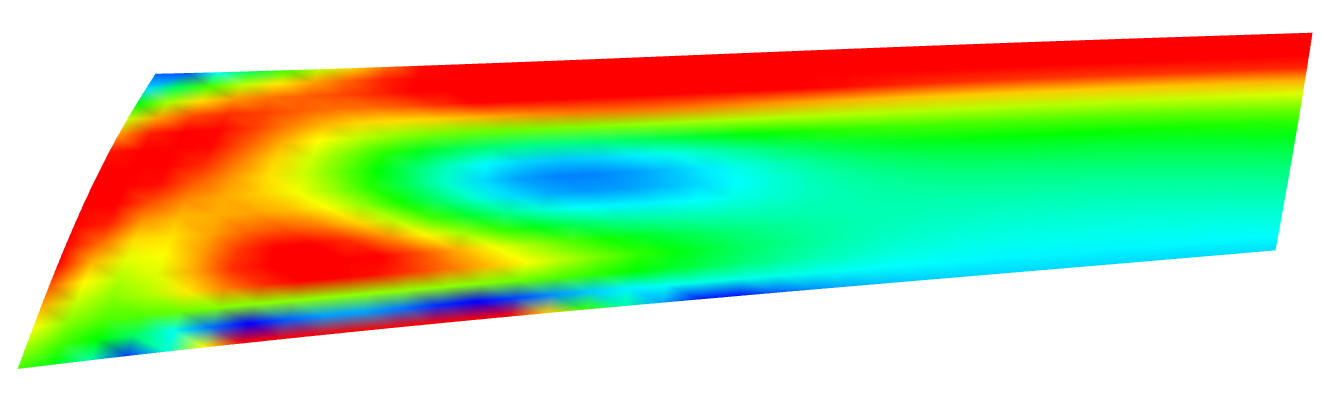}
		\end{minipage}
	}
	\subfigure[]{
		\begin{minipage}[b]{.45\linewidth}
			\centering
			\includegraphics[scale=0.1]{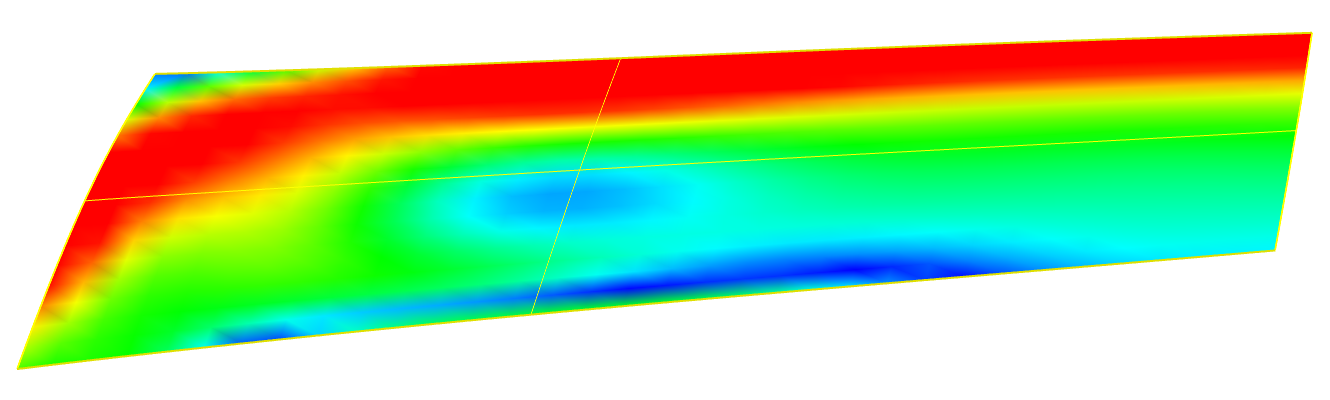}
		\end{minipage}
	}
	\subfigure[]{
		\begin{minipage}[b]{.45\linewidth}
			\centering
			\includegraphics[scale=0.1]{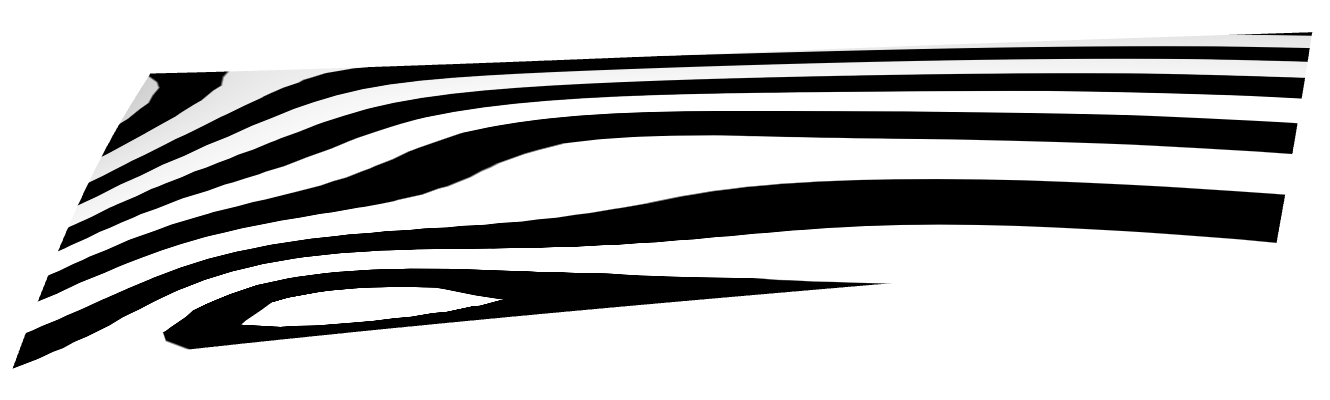}
		\end{minipage}
	}
	\subfigure[]{
		\begin{minipage}[b]{.45\linewidth}
			\centering
			\includegraphics[scale=0.1]{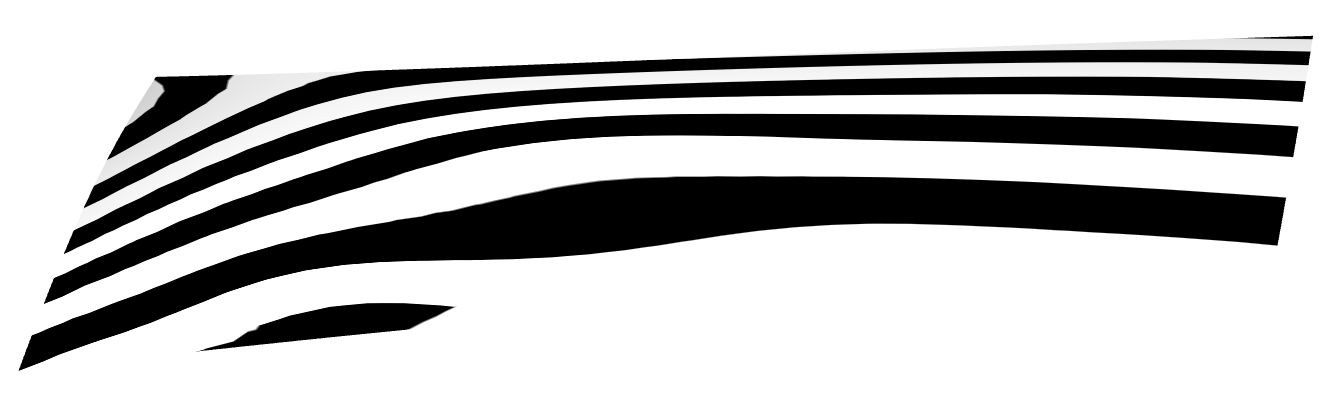}
		\end{minipage}
	}
	\caption{\label{fig11}Fairing results for the surface model. (a)(c) Original model. (b)(d) The  surface fairing by our method. (a)(b) Absolute value of mean curvature on the surfaces.  (c)(d) The zebra diagrams on the surfaces.}
\end{figure}

Figure \ref{fig11} is a surface with $8\times 13$ control points, derived from one of the surfaces in Figs.\ref{fig8}(a). Figs.\ref{fig11}(a) shows the mean curvature map of the original surface, and Figs.\ref{fig11}(b) shows its zebra stripe. We select to adjust the number of control points to 12 and set the fairing weight to 0.0005. Figs.\ref{fig11}(c) and \ref{fig11}(d) present the results after fairing.

\section{Conclusion}

A curve and surface fairing method based on progressive-iterative approximation is presented, along with a proof of its convergence. Each control point is assigned an independent weight, enabling precise shape control. The traditional energy minimization approach is shown to be a special case of the proposed method, which further supports its capacity to achieve fairness with minimal perturbation of the original geometry. The method can fair the curve surface globally, and can also adjust the shape of the curve surface more finely through many parameters to achieve fairness improvement or local feature preservation. Numerical experiments demonstrate the effectiveness and efficiency of the algorithm. Additionally, an automatic control point selection strategy based on the order of energy difference is introduced for efficient adjustment.



\end{document}